\documentclass[leqno]{article}                
\usepackage{amssymb}
\newcommand{\beq}{\begin{equation}}           
\newcommand{\eeq}{\end{equation}}            
\newcommand{\ba}{\begin{array}}
\newcommand{\ea}{\end{array}}

\newcommand{\sect}[1]{\setcounter{equation}{0}\section{#1}}

\newcommand{\bea}{\begin{eqnarray}}
\newcommand{\eea}{\end{eqnarray}}
        
\newtheorem{lemma}{Lemma}
\newtheorem{theorem}{Theorem}

\def\sq{\mbox{\rlap{$\sqcap$}$\sqcup$}}
\newenvironment{proof}[1]{\vspace{5pt}\noindent{\sc Proof #1}\hspace{6pt}}%
{\hfill\sq}
\newcommand{\bp}{\begin{proof}}
\newcommand{\ep}{\end{proof}\par\vspace{10pt}\noindent}
\def\b#1{{\mathbb #1}}

\def\nn{\nonumber  \\}

\date{}

\begin{document}

\title{\bf Global Stability properties for a class of dissipative 
           phenomena via one or several Liapunov functionals}
 \author{ {\sc  A. D'Anna   \hspace{30mm} G. Fiore}  \\\\
  Dip. di Matematica e Applicazioni, Fac.  di Ingegneria\\
        Universit\`a di Napoli, V. Claudio 21, 80125 Napoli
       }
 \maketitle

{\sc Riassunto}: {\small\it Si determinano nuovi risultati
sull'esistenza, unicit\`a, limitatezza, 
stabilit\`a ed attrattivit\`a delle soluzioni di una classe
di problemi di
valori iniziali e al contorno caratterizzati da un'equazione 
quasi lineare del terzo ordine che pu\'o avere 
un termine forzante non autonomo.
La classe include equazioni che si incontrano in Teoria della
Superconduttivit\`a, Meccanica Quantistica e Teoria dei 
Materiali Viscoelastici.}

 \vspace{1mm}

{\sc Abstract}: {\small\it We find some new results regarding the 
existence,
uniqueness, boundedness, stability and attractivity of the solutions
of a class of initial-boundary-value problems characterized
by a quasi-linear third order equation which may 
have non-autonomous forcing terms.
The class includes equations arising in Superconductor Theory,
Quantum Mechanics and in the Theory of Viscoelastic Materials.}

 \vspace{1mm}


 \vspace{1mm}

 {\sc A.M.S. Classification}: {\small 35B35 - 35G30}

\bigskip
\noindent
Preprint 02-38 Dip. Matematica e Applicazioni, Universit\`a di Napoli

\sect{Introduction}

In this paper we deal with questions regarding the existence,
uniqueness, boundedness, stability and attractivity of solutions $u$
of the following class of initial-boundary-value problems:
 \beq                                                 \label{21}
  \hspace*{5mm}
Lu=f(x,t,u,u_x,u_{xx},u_t), \qquad  \qquad 0< x < 1, \quad 0 < t < T,
 \eeq
 where $L=-\varepsilon \partial_{xxt}-c^2\partial_{xx}+\partial_{tt}$,
$f$ is a continuous function of its arguments,
$c$ and $\varepsilon$ are positive constants, and
 \beq                                                   \label{22}
   u(x,0)=u_0(x),~~u_t(x,0)=u_1(x),  ~~0 < x < 1,
 \eeq
 \beq                                                   \label{23}
   u(0,t)=h_1(t),~~u(1,t)=h_2(t),  ~~0 < t < T,
 \eeq
where $T\le +\infty$,
$h_1,h_2\in C^2([0,T[)$, $u_0,u_1\in C^2([0,1])$ 
are assigned and fulfill the consistency condition
\beq
h_1(0)=u_0(0), \quad \dot h_1(0)=u_1(0) \qquad    
h_2(0)=u_0(1), \quad \dot h_2(0)=u_1(1) .       \label{23bis}
\eeq
Solutions $u$ of such
problems describe a number of physically remarkable
continuous phenomena occurring on a finite space interval. 
In the operator $L$ the D'Alembert\-ian 
$-c^2\partial_{xx}+\partial_{tt}$ induces wave propagation,
$-\varepsilon \partial_{xxt}$ dissipation. The term
at the right-hand side of (\ref{21}) may contain
forcing terms, nonlinear (local) couplings of $u$ to
itself, further dissipative terms. For instance,
when $f=-b \sin u-au_t+ F(x,t)$, where
$a,b$ are positive constants, we deal with the perturbed
Sine-Gordon  equation, which can be used e.g.
to describe the classical Josephson effect  with driving force $F$
in the Theory of Superconductors \cite{dav,lsc}. 
$F$ is a forcing term, $-au_t$ a dissipative one and $-b \sin u$
a nonlinear coupling. On the other hand
it is well known \cite{mor} that equation (\ref{21})  describes
the evolution of the displacement $u(x,t)$ of the section of a rod
from its rest position $x$ in a
Voigt material when an external force $f$ is applied; 
in this case $c^2=E/\rho$, $\varepsilon=1/(\rho\mu)$,
where $\rho$ is the (constant) linear density of the rod at rest,
and $E,\mu$ are respectively
the elastic and viscous constants of the rod, which enter
the stress-strain relation
$\sigma=E\nu+\partial_t \nu/\mu$,
where $\sigma$ is the stress, $\nu$ is the strain.
As we shall see in the sequel,
even cosidering only one of these examples, e.g. the perturbed
Sine-Gordon equation $f=-b \sin u-au_t$, it is important 
to keep room for a more
general $f$ because the latter will naturally appear
when asking whether a particular solution $u^*$ of the
problem is stable or attractive, or when reducing the original
problem to one with trivial boundary conditions.

Several papers \cite{DacDan98,dr1,dr2,rio,ghi,gr1,gr2,DanFio00}
have already been devoted to the analysis of the operator $L$
and more specifically to the investigation
of the boundedness, stability and attractivity 
of the solutions of the above problem. Here we improve
previous results, by weakening the assumptions on $f$,
and find some new ones. In Section \ref{existunique} we 
improve the existence and uniqueness Theorem 2.1 proved in 
\cite{DacDan98}, in that we
require $f$ to satisfy only {\it locally}
a Lipschitz condition. In Section \ref{boundstab} we improve 
the boundedness and stability Theorem 3.1 of the same reference,  
in that we require only a suitable {\it time average}
of the quadratic norm of $f$ to be bounded. While doing so
we prove two lemmas concerning boundedness and attractivity 
of the null solution for a class of first order ordinary differential 
equations in one unknown; the second lemma is a generalization of
a lemma due to Hale \cite{Hale}.
In Sections \ref{expstab} and \ref{nonan} we respectively
improve the exponential asymptotical stability Theorem 3.3 
of   \cite{DacDan98} and the uniform asymptotical stability Theorem 
2 of  \cite{DanFio00}, valid for some special $f$, by removing the
boundedness assumption on the latter.
The trick we use is to associate to each neighbourhood of the origin 
with radius $\sigma$ (the `error') a Liapunov functional
depending on a parameter $\gamma$  adapted to $\sigma$,  instead
of fixing $\gamma$ once and for all.

\section{Existence and uniqueness of the solution}
\label{existunique}
 \setcounter{equation}{0}

To discuss the existence and uniqueness of the above problem
it is convenient to formulate it as an equivalent integro-differential
equation so as to apply the fixed-point theorem. 

As in \cite{DacDan98}, we start from the identity
 \beq                                                   \label{24}
    \partial_\xi
   (c^2uw_\xi-c^2u_\xi w+\varepsilon u_\xi w_\tau-\varepsilon uw_{\xi\tau})
   +\partial_\tau(u_\tau w-uw_\tau-\varepsilon u_{\xi\xi}w)=
 \eeq
 \[
   =fw-u(\varepsilon w_{\xi\xi\tau}-c^2w_{\xi\xi}+w_{\tau\tau}),
 \]
 that follows from (\ref{21}) for any smooth function
 $w(\xi,\tau)$, assuming $u(\xi,\tau)$ is a smooth solution of
 (\ref{21}). We choose $w$ as a function depending also
 on $x,t$ and fulfilling the equation $Lw=0$, more precisely
 \beq                                                   \label{25}
  w(x,\xi,t-\tau)=\theta(x-\xi,t-\tau)-\theta(x+\xi,t-\tau),
 \eeq
 with
 \beq                                                        \label{26}
   \theta(x,t) =
   K(|x|,t)+\sum_{m=1}^{\infty}\left[K(|x+2m|,t)+K(|x-2m|,t)\right].
 \eeq
 The function $K$ represents the fundamental solution of the 
 linear equation $LK=0$.
 It has been determined and studied in \cite{dr1}, and reads
  \beq \hspace*{4mm}                                               \label{27}
  K(|x|,t)=\int_0^t  \frac{e^{-c^2\tau/\varepsilon}}{\sqrt{\pi\varepsilon\tau}}
    d\tau\int_0^\infty\frac{x^2(z+1)}{4\varepsilon\tau}
    e^{-x^2(z+1)^2/4\varepsilon\tau}
    I_0\left(\frac{c}{\varepsilon}2|x|\sqrt{z}\right)dz,
 \eeq
 where $I_0$ is the modified Bessel function of order zero.
 Since $\theta(-x,t) = \theta(x,t)$ and
 $\theta(x+2m,t) = \theta(x,t), m\in N$, it is sufficient
 to restrit our attention to the domain $0\leq x<2$, and note that
 $\theta$ is continuous together its partial derivatives and satisfies the
 equation $L\theta=0.$ Moreover, from the analysis of $K$ developed in \cite{dr1},
 we can deduce that $\theta$ is a positive function that has properties
 similar to ones of the analogous function $\theta$ used for the heat operator,
 see \cite{cannon}.

{\bigskip

As for the data we shall assume that:
 \beq                                                           \label{28}
 \mbox{$f(x,t,n,p,q,r)$ is defined and continuous on the set}
 \eeq
 \[
 \{(x,t,n,p,q,r)~|~ 0\leq x \leq 1,~0\leq t \leq T,~-\infty<n,p,q,r<\infty,
 \  \ T>0 \},
 \]
\bea                                                           \label{29}
&&\mbox{ it locally satisfies a Lipschitz condition, namely for any} \\
&&\mbox{ bounded set $\Omega\subset[0,T]\times\b{R}^4$ 
there exists a constant $\mu_{\Omega}$} \nn
&&\mbox{ such that for any 
$(t,n_1,p_1,q_1,r_1), (t,n_2,p_2,q_2,r_2)\in\Omega$}\nn
&& \mbox{and}~x\in[0,1]\nonumber 
 \eea
 \[
    |f(x,t,n_1, p_1,q_1,r_1)-f(x,t,n_2,p_2,q_2,r_2)|
   \leq
 \]
 \[
   \leq \mu_{\Omega}\{|n_1-n_2|+|p_1-p_2|+|q_1-q_2|+|r_1-r_2| \},
 \]
  \beq                                                             \label{210}
  u_0,~u'_0,~u_0^{''},~u_1~ \mbox{continuous on}~  0\leq x\leq 1,
 \eeq
 \beq                                                             \label{211}
  h_i,~\dot h_i,~i=1,2,~ \mbox{continuous on} ~ 0\leq t\leq T,
 \eeq
 \beq                                                             \label{212}
  h_1(0)=u_0(0),~h_2(0)=u_0(1),~\dot h_1(0)=u_1(0),~\dot h_2(0)=u_1(1).
 \eeq

Given a solution $u$ of (\ref{21})-(\ref{23}), by integrating (\ref{24}) on 
$\{(\xi,\tau) \ | \ 0<\xi<1, \delta<\tau<t-\delta\}$,
$\delta>0$ and letting $\delta\rightarrow 0$,
we find that it satisfies the following integral equation  
 \beq                                                 \label{213}
  u(x,t)=  \int_0^1 w_t(x,\xi,t)
  u_0(\xi)d\xi+  \int_0^1 w(x,\xi,t) [u_1(\xi)-\varepsilon u_0^{''}(\xi)]d\xi-
 \eeq
 \[
    -2\int_0^th_1(\tau)(c^2+\varepsilon\partial_t)\theta_x(x,t-\tau)d\tau+
     2\int_0^th_2(\tau)(c^2+\varepsilon\partial_t)\theta_x(1-x,t-\tau)d\tau+
 \]
 \[
 +\int_0^td\tau\int_0^1w(x,\xi,t-\tau)
  f(\xi,\tau,u(\xi,\tau),u_\xi(\xi,\tau),u_\tau(\xi,\tau),u_{\xi\xi}(\xi,\tau))
  d\xi.
 \]

Conversely,  one can immediately verify that under the assumptions
(\ref{28})-(\ref{212}) a solution $u$ of (\ref{213})
satisfies (\ref{21}) using the fact that $L\theta=0$ and $Lw=0$. 
We refer the reader to  \cite{DacDan98} for
the slightly longer proof that the initial conditions
(\ref{22}) and the boundary conditions are satisfied.

 If $f=f(x,t)$, (\ref{213}) gives the unique explicit
 solution of (\ref{21})-(\ref{23}). On the contrary, if $f$ depends
 on $u$  (\ref{213}) is an integro-differential equation.
 We shall now discuss the existence and uniqueness 
 of its solutions. 
 
For any $c,d\in[0,T]$, $c\le d$ we shall denote
\[
{\cal B}_{[c,d]}:=\{u(x,t)~|~ u,u_x,u_t,u_{xx}\in C([0,1]\times[c,d]), \}.
\]
For any $a\in[0,T]$, $v\in{\cal B}_{[0,a]}$
and $t\in [a,T]$ we define a mapping of ${\cal B}_{[a,T]}$ into
itself by
 \bea                                                 \label{214}
 \hspace*{8mm}
&&  {\cal T}_vu(x,t) :=  \omega_v(x,t)\\
  &&+\int_a^td\tau\int_0^1w(x,\xi,t-\tau)
  f(\xi,\tau,u(\xi,\tau),u_\xi(\xi,\tau),u_\tau(\xi,\tau),u_{\xi\xi}(\xi,\tau))
  d\xi \nonumber
 \eea
 where
 \bea                                                 \label{defomega}
&& \omega_v(x,t) =\int_0^1 w_t(x,\xi,t)
  u_0(\xi)d\xi+  \int_0^1 w(x,\xi,t) [u_1(\xi)-
  \varepsilon u_0^{''}(\xi)]d\xi- \nn
&&  -2\int_0^th_1(\tau)(c^2+\varepsilon\partial_t)\theta_x(x,t-\tau)d\tau
  + 2\int_0^th_2(\tau)(c^2+\varepsilon\partial_t)\theta_x(1-x,t-\tau)d\tau\nn
  &&+\int_0^ad\tau\int_0^1w(x,\xi,t-\tau)
  f(\xi,\tau,v(\xi,\tau),v_\xi(\xi,\tau),v_\tau(\xi,\tau),v_{\xi\xi}(\xi,\tau))
  d\xi. \nonumber
\eea 
We fix a $\rho>0$ and for any $t\in[a,T]$ we consider the domain
\bea
&& S_{v,t}:=\{u\in{\cal B}_{[a,T]} \quad ~|~ 
\quad\forall x\in[0,1]\:\:
|u(x,t)-\omega_v(x,t)|<\rho, \qquad   \nn
&&\:|u_x(x,t)\!-\!\omega_v{}_x(x,t)|<\rho,
\:|u_{xx}(x,t)\!-\!\omega_v{}_{xx}(x,t)|<\rho,
\:|u_t(x,t)\!-\!\omega_v{}_t(x,t)|<\rho\} \nonumber
\eea
and define
\bea 
&&M=M(a,T,v,\rho):=\sup_{\tau\in[a,T] \atop
\xi\in[0,1]}\sup_{u\in S_{v,\tau}}
| f(\xi,\tau,u(\xi,\tau),u_\xi(\xi,\tau),
u_\tau(\xi,\tau),u_{\xi\xi}(\xi,\tau))|\nn
\label{defb}&&\qquad\qquad \qquad 
b-a=\min\left\{T-a,\frac{\rho}{M},\frac{c\rho}{M},
\frac{\varepsilon\rho}M,\sqrt{\frac{2\rho}{M}}\right\}\\
&&R_{a,b,v}:=\{u\in{\cal B}_{[a,b]} \quad~|~ \quad
\forall (x,t)\in[0,1]\times[a,b] \:\:
|u(x,t)\!-\!\omega_v(x,t)|\le\rho, \nn
&& |u_x(x,t)\!-\!\omega_v{}_x(x,t)|\le\rho,
\:|u_{xx}(x,t)\!-\!\omega_v{}_{xx}(x,t)|\le\rho,
\:|u_t(x,t)\!-\!\omega_v{}_t(x,t)|\le\rho\}.
\nonumber
\eea
Note that by its definition $M$ is finite because $f$ is 
continuous and evaluated on a compact subset of $\b{R}^6$.
We denote by $\mu=\mu(a,b,v,\rho)$ the constant $\mu_{\Omega}$ of (\ref{29})
corresponding to the choice 
\bea
&&\Omega=\{(t,n,p,q,r)~|~\mbox{with }t\in[a,b], \mbox{ and such that }
\exists x\in[0,1],\,\exists u\in R_{a,b,v} \nn
&&\qquad \mbox{ such that }
n=u(x,t),\,p=u_x(x,t),\,q=u_{xx}(x,t),\,r=u_t(x,t)\}, \nonumber
\eea
we choose a positive constant $\lambda$
\beq                                                 \label{215}
  \lambda=\lambda(a,b,v,\rho)> \max 
  \left\{1, \mu\left(2+\frac 1c +\frac{1+2c^2}
  {\varepsilon}\right) \right\}
\eeq
and we introduce a norm 
\beq
\|u\|_{a,b}:=\sup_{[0,1]\times[a,b]}|
{\rm e}^{-\lambda t}u(x,t)|+
\sup_{[0,1]\times[a,b]}|{\rm e}^{-\lambda t}u_x(x,t)|+
\eeq
\[
+\sup_{[0,1]\times[a,b]}|{\rm e}^{-\lambda t}u_t(x,t)|
+\sup_{[0,1]\times[a,b]}|{\rm e}^{-\lambda t}u_{xx}(x,t)|.
 \] 
We now show that ${\cal T}_v$ is a map of $R_{a,b,v}$
into itself, more precisely a contraction 
(w.r.t the above norm).
From (\ref{214}) we get for any $(x,t)\in[0,1]\times[a,b]$
 \[
 |{\cal T}_vu(x,t)-\omega_v(x,t)| \leq M(a,T,v,\rho)
 \int_a^td\tau\int_0^1 |w(x,\xi,t-\tau)|d\xi,
 \]
 and, because of the inequality \cite{dr1}
 \beq
 \qquad  \int_0^1 |w(x,\xi,t-\tau)| d\xi= \int_0^1
   |\theta(x-\xi,t-\tau)-\theta(x+\xi,t-\tau)| d\xi \leq t-\tau,
   \label{ineq1}
 \eeq
 and the definition of $b$  we find
\beq
 |{\cal T}_vu(x,t)-\omega_v(x,t)| \leq M(a,T,v,\rho)
\frac{(b-a)^2}2 \le\rho.
\eeq
Similarly, one can prove that
\bea
 &&|{\cal T}_vu_x(x,t)-\omega_v{}_x(x,t)| \leq \rho, \\
 &&|{\cal T}_vu_{xx}(x,t)-\omega_{vxx}(x,t)| \leq \rho, \\
 &&|{\cal T}_vu_t(x,t)-\omega_v{}_t(x,t)| \leq \rho, 
\eea
making use of the basic properties of $K$ proved in \cite{dr1},
which lead to the following estimates:
 \bea
&& \int_0^1 |w_x(x,\xi,t-\tau)| d\xi \leq 1/c, \nn 
&&   \int_0^1 |w_t(x,\xi,t-\tau)| d\xi \leq 1,   \label{ineq2}\\
&&  \int_0^1 |w_{xx}(x,\xi,t-\tau)| d\xi  \leq 
   \frac 1{\epsilon}[1+ 2c^2(t-\tau)].            \nonumber
 \eea
The first two inequalities were already given in \cite{DacDan98}, 
together with 
\beq                                               \label{ineq3}
 \int_0^1 |(\partial_t-\partial_x^2) w(x,\xi,t-\tau)| d\xi \leq 1.
\eeq
The third was used but not explicitly written, and
easily follows from the latter inequality, the equation $L\theta=0$,
the relation $\theta(x,0)=0$. In fact, from $L\theta=0$ it immediately
follows that
$$
\theta_t-\theta_{xx}=\partial_t[\theta+\frac{\epsilon}{c^2} \theta_{xx}-
\frac 1{c^2}\theta_t],
$$
and therefore
$$
w_t(x,\xi,t-\tau)-w_{xx}(x,\xi,t-\tau)=\partial_t[w(x,\xi,t-\tau)+
\frac{\epsilon}{c^2} w_{xx}(x,\xi,t-\tau)-
\frac 1{c^2}w_t(x,\xi,t-\tau)];
$$
Integrating over $\xi$ and using (\ref{ineq3}) we find
$|\partial_t A(x,t-\tau)| \le 1$, where 
$$ 
A(x,t-\tau):= \int_0^1 [w(x,\xi,t-\tau)+
\frac{\epsilon}{c^2} w_{xx}(x,\xi,t-\tau)-
\frac 1{c^2}w_t(x,\xi,t-\tau)]d\xi.
$$
As $\theta(x,0)=0$, then
$A(x,0)=0$. By the comparison principle we therefore
find 
$$
\tau-t\le  A(x,t-\tau)= \int_0^1 wd\xi+\int_0^1\frac{\epsilon}{c^2} w_{xx} 
d\xi-\int_0^1 \frac 1{c^2}w_td\xi \le t-\tau,
$$ 
implying 
$$
\left|\int_0^1\frac{\epsilon}{c^2} w_{xx} \right|\le  (t-\tau)+
\left|\int_0^1 wd\xi\right|+\left|\int_0^1 \frac 1{c^2}w_td\xi \right|;
$$
using (\ref{ineq1}) and  (\ref{ineq2})$_2$ to bound the
integrals at the right hand-side we find (\ref{ineq2})$_3$.

From the above results we conclude that 
 ${\cal T}_vu(x,t)\in R_{a,b,v}$ as claimed.

From (\ref{214}), (\ref{ineq1}) we get for $t\in[a,b]$
 \bea
&&|{\cal T}_vu_1(x,t)-{\cal T}_vu_2(x,t)|{\rm e}^{-\lambda t}\leq 
\mu \|u_1\!-\!u_2\|_{a,b}
 \int_a^t{\rm e}^{-\lambda (t-\tau)}d\tau\int_0^1\! |w(x,\xi,t\!-\!\tau)|d\xi\nn
 \label{216}
 && \qquad \qquad \le\mu\|u_1-u_2\|_{a,b}
 \int_a^t{\rm e}^{-\lambda (t-\tau)}(t-\tau)d\tau
 \leq \frac{\mu}{\lambda^2}\|u_1-u_2\|_{a,b}.
 \eea
 
From (\ref{214}), (\ref{ineq2}) 
 one can get analogous results for the partial derivatives
 $\partial_x$, $\partial_t$, $\partial_x^2$ of (\ref{214}):
\bea
&&|{\cal T}_vu_{1x}(x,t)-{\cal T}_vu_{2x}(x,t)|{\rm e}^{-\lambda t}\leq 
 \frac{\mu}{\lambda c}\|u_1-u_2\|_{a,b},\nn
&&|{\cal T}_vu_{1t}(x,t)-{\cal T}_vu_{2t}(x,t)|{\rm e}^{-\lambda t}\leq 
 \frac{\mu}{\lambda}\|u_1-u_2\|_{a,b},\\
&&|{\cal T}_vu_{1xx}(x,t)-{\cal T}_vu_{2xx}(x,t)|{\rm e}^{-\lambda t}\leq 
 \frac{\mu}{\lambda\varepsilon}(1+\frac{2c^2}{\lambda})\|u_1-u_2\|_{a,b}.
 \nonumber
\eea
Thus, we  obtain
 \beq                                                       \label{217}
 \qquad\|{\cal T}_vu_1(x,t)-{\cal T}_vu_2(x,t)\|_{a,b}\leq  \frac{\mu}
 {\lambda}\left[\frac{1}{\lambda}+
 \frac{1}{c}+1
   +\frac 1{\varepsilon}+\frac{2c^2}{\varepsilon\lambda} \right] 
   \|u_1-u_2\|_{a,b},
 \eeq
 with $\mu\equiv\mu(a,b,v,\rho)$, $\lambda\equiv\lambda(a,b,v,\rho)$.
 Under assumption (\ref{215}), inequality (\ref{217}) shows that 
 ${\cal T}_v$ is
 a contraction of $R_{a,b,v}$ into itself. Thus  we are in the
 conditions to apply the fixed point theorem, and we find that
 there exists a unique solution in $R_{a,b,v}$  
 of the problem ${\cal T}_vu=u$ in the time interval $[a,b]$.

 We now apply the above result iteratively. We start by choosing
 $a=0=:a_0$, $v=0$; the last integral disappears from (\ref{defomega}).
 From  the definition of $b$ we determine the corresponding $b=:a_1$
 and by the fixed point theorem a unique solution
 $u^{(1)}(x,t)$ of the problem (\ref{21})-(\ref{23bis}) 
  in the time interval $[0,a_1]$. Next we choose
 $a=a_1$, $v=u^{(1)}$; from  (\ref{defb}) we determine the 
 corresponding $b=:a_2$
 and by the fixed point theorem a unique solution
 of the problem ${\cal T}_{u^{(1)}}u=u$
 in the time interval $[a_1,a_2]$. This is also a smooth
 continuation of $u^{(1)}$, therefore we have found
 a unique solution  $u^{(2)}(x,t)$ of the problem (\ref{21})-(\ref{23bis}) 
 in the time interval $[0,a_2]$, and so on. 
 We conclude by stating the following

 \vspace{3mm}

 {\bf Theorem} {\it Under hypotheses} (\ref{28})-(\ref{212}), {\it the
 quasilinear problem} (\ref{21})-(\ref{23}) {\it has a unique smooth solution
 in the time interval $[0,a_{\infty}]$,} 
 where 
 \[
 a_{\infty}:=\lim_{k\to+\infty}a_k\leq T.
 \]

\sect{Eventual boundedness and asymptotic stability}
\subsection{Preliminaries}

By the rescaling $t\to  t/c$ $\varepsilon\to c\varepsilon$ and
of $f\to c^2 f$ we can
factor $c$ out  of (\ref{21}), so that it completely disappears from 
the problem, without loosing generality. In the sequel 
we shall assume we have done this.
Moreover, without loss of generality  we can also
consider $h_1(t)=h_2(t)\equiv 0$ in (\ref{23}),
as any problem (\ref{21}-\ref{23bis}) is equivalent to
another one of the same kind with trivial
boundary conditions and a different $f$.
In fact, setting for any $t\in J:=[0, \infty[$
$$
v(x,t):=u(x,t)+p(x,t)\qquad\qquad p(x,t):=(1-x)h_1(t)+
xh_2(t)
$$
we immediately find that $v(0,t)=v(1,t)\equiv 0$, that the initial
condition for $v,v_t$ are completely determined
and that $v$ fulfills the equation
$$
-\varepsilon v_{xxt}+v_{tt}-v_{xx}=
\tilde f(x,t,v,v_x,v_{xx},v_t),
$$
where
$$
\tilde f(x,t,v,v_x,v_{xx},v_t):=
f(x,t,v-p,v_x-h_2+h_1,v_{xx},v_t-p_t)-
p_{tt}.
$$
The difference $u:=\tilde u-u^*$ 
between a generic solution $\tilde u$ and a given one
$u^*$ of the problem
(\ref{21}-\ref{23bis}) is also a solution of a new problem of
the same kind, which we denote by problem ${\cal P}$,
but with $h_1(t)\equiv h_2(t)\equiv 0$, namely
\beq                                                 \label{31}
\ba{l}
\qquad -\varepsilon u_{xxt}+u_{tt}-u_{xx}=
f(x\!,\!t\!,\!u\!,\!u_x,\!u_{xx},\!u_t),
\qquad x\!\in]0,1[, \:\: t>t_0 \in J \\[6pt]
\qquad u(0,t)=0,~~u(1,t)=0,  \qquad \qquad \qquad\qquad \qquad  t \in J   
\ea  
\eeq
with the initial conditions
 \beq                                                   \label{32}
   u(x,0)=u_0(x),~~u_t(x,0)=u_1(x),\qquad   x\!\in]0,1[,
 \eeq
fulfilling the consistency conditions
\beq
u_0(0)=u_1(0)=u_0(1)=u_1(1)=0.                 \label{32bis}
\eeq
Here 
$$
f(x\!,\!t\!,\!u\!,\!u_x,\!u_{xx},\!u_t)=
f(x\!,\!t\!,\!u\!+\!u^*,\!u_x\!+\!u^*_x,\!u_{xx}\!+
\!u^*_{xx},\!u_t\!+\!u^*_{xx})-
f(x\!,\!t\!,\!u^*\!,\!u^*_x,\!u^*_{xx},\!u^*_t)
$$
and $u_0(x):=\tilde u_0(x)-u^*_0(x)$, $u_1(x):=\tilde u_1(x)-u^*_1(x)$.
The two solutions $\tilde u,u^*$ are `close' to each other iff $u$
is a `small' solution of the latter problem, and coincide iff $u$
is the null solution.

We introduce
the distance between the origin $O$ and a nonnull element 
$\big(u(\cdot,t),u_t(\cdot,t)\big)\in 
\Gamma:=\big(C_0([0, 1])\cap C^2([0, 1])\big)\times C_0([0, 1])$ 
as the functional $d(u,u_t)$, where for any
$(\varphi,\psi)\in\Gamma$
we define
 \beq                                                \label{33}
  d^2(\varphi,\psi) =
  \int_0^1(\varphi^2+ \varphi_x^2+ \varphi^2_{xx} + \psi^2)dx.
 \eeq
The notions of (eventual) boundedness,
stability, attractivity, etc.  
are formulated using this distance. Imposing the condition
that $\varphi,\psi$ vanish in $0,1$ one easily derives that
$|\varphi(x)|,|\varphi_x(x)|\le d(\varphi,\psi) $ for any $x$;
therefore a convergence in the norm $d$ implies also a
uniform pointwise convergence of  $\varphi,\varphi_x$.

\bigskip

{\sc Definition 3.1}. The solutions of $(\ref{31}-\ref{32bis})$ are  eventually
uniformly bounded  if for any $\alpha>0$ there exist an  $s(\alpha)\geq 0$
and a $\beta(\alpha)>0$ such that if  $t_0\geq s(\alpha)$, 
$d(u_0,u_1)\leq \alpha$,
then $d(u(t),u_t(t))<\beta(\alpha)$ for all $t\geq t_0$.  If $s(\alpha)=0$ the
solutions of (\ref{31}) are uniformly bounded.

\bigskip

 {\sc Definition 3.2}. The origin $O$ of $\Gamma$ is eventually
quasi-uniform-asymptotically stable in the large 
for the solutions of $(\ref{31})$ if
for any $\rho,\alpha>0$ there exist $s(\alpha)\ge 0$, and 
$\hat T(\rho,\alpha)> 0$
such that if $d(u_0,u_1)\le\alpha$, 
$t_0\ge s(\alpha)$ then $d(u,u_t)<\rho$ for any
$t\ge t_0+\hat T$. If $s(\alpha)=0$, $u(x,t)\equiv 0$ is said to be
quasi-uniform-asymptotically stable in the large  for the
solutions of $(\ref{31})$.

\bigskip

Suppose now that problem  ${\cal P}$
admits the solution $u(x,t)\equiv 0$.

{\sc Definition 3.3}. The solution $u(x,t)\equiv 0$ is
uniform-asymptotical stable in the large if it is uniformly stable
and quasi-uniform-asymptotically stable in the large, and
the solutions of problem  ${\cal P}$
are uniformly bounded. 

\bigskip

 {\sc Definition 3.4}. The solution $u(x,t)\equiv 0$ of the problem
${\cal P}$
 is exponential-asymptotically stable in the large if
for any $\alpha>0$ there are two positive constants
$D(\alpha), C(\alpha)$ such that if $d(u_0,u_1)\leq \alpha$, then
\beq
   d(u(t),u_t(t))\leq D(\alpha)
   \exp \left[-C(\alpha)(t-t_0) \right] d(u_0,u_1), \quad   \forall
   t\geq t_0.                                     \label{range}
\eeq

\bigskip

To prove our theorems we shall use the Liapunov direct method.
We introduce the Liapunov functional
 \beq                                                 \label{34}
     V(\varphi,\psi)=\frac{1}{2}\int_0^1\{(\varepsilon \varphi_{xx}-\psi)^2+
     \gamma\psi^2+(1+\gamma)\varphi_x^2 \}dx,
 \eeq
 where $\gamma$ is an arbitrary positive constant. 
Using the inequality $|2\varepsilon \varphi_{xx}\psi|\le 
\varepsilon(\varphi_{xx}^2+\psi^2)$ we find
\[
  V\leq \frac{1}{2}\int_0^1\{\varepsilon^2 \varphi_{xx}^2 +\psi^2
   +\varepsilon \varphi_{xx}^2 +\varepsilon\psi^2
    + \gamma\psi^2+(1+\gamma)\varphi_x^2 \}dx.
 \]
 Setting
 \beq                                                 \label{35}
     c_2^2= \max \{\varepsilon(1+\varepsilon)/2,
           (1+\varepsilon+\gamma)/2\},
 \eeq
 we thus derive
 \beq                                                 \label{36}
     V(\varphi,\psi)\leq c^2_2d^2(\varphi,\psi).
 \eeq
Moreover, it is known that \cite{rio}
\beq                                       \label{38}
  \varphi(0)=0, ~~\varphi(1)=0 ~~ \Longrightarrow ~~
\left\{
\ba{l} 
\int_0^1 \varphi^2_x(x)dx\geq \pi^2\int_0^1 \varphi^2(x)dx\\
\qquad\\
\int_0^1 \varphi^2_{xx}(x)dx\geq \pi^2\int_0^1 \varphi_x^2(x)dx
\ea\right.
 \eeq
(these inequalities can be easily proved by Fourier
analysis of $\varphi$). In view of the bounds we shall
consider below we introduce the notation
\beq
\qquad\omega_1\!:=\!\frac{\pi^4}{1\!+\!\pi^4}\!\approx\! 0.99\qquad
\omega_2\!:=\!\frac{\pi^4}{1\!+\!\pi^2\!+\!\pi^4}\!\approx\! 0.90,\qquad
\omega_3\!:=\!\frac{\pi^2}{1\!+\!\pi^2}\!\approx\! 0.91.
\eeq

 Using (\ref{38}) and an argument employed in \cite{DacDan98}, we
 get
  \beq                                       \label{39}
     V(\varphi,\psi)\geq c^2_1d^2(\varphi,\psi).
 \eeq
 where
 \vspace{2mm}
 \beq                                       \label{310}
 c_1^2= \min \{\frac{\varepsilon^2}8
 \omega_1, \frac 12\left(\gamma-\frac 12\right)\},
        \qquad (\gamma >1/2).
 \eeq
Therefore, from (\ref{36}) and (\ref{39}) we find
\beq
\frac V{c_2^2} \le d^2 \le \frac V{c_1^2}.                \label{invemag}
\eeq
On the other hand, choosing $\gamma=1$ in (\ref{34}) 
and reasoning as it has been done in  \cite{DacDan98}
it turns out that
\bea                                         \label{inter}
 \dot V &=& \int_0^1 \left \{
    -\frac{\varepsilon }{2}u_{xx}^2-\varepsilon u_{xt}^2+
    \frac{\varepsilon}{2}u_{t}^2
       -\frac{\varepsilon}{2}   \left(u_{xx}+f\right)^2 
 \right. \nn
&& \left. \qquad\qquad
  -\frac{\varepsilon}{2}\left(u_t-2f/\varepsilon \right)^2+
  Af^2 \right \}dx \nn
&\le & -\int_0^1 \left\{\frac{\varepsilon }2\omega_2 \left(u^2+u_x^2+
u_{xx}^2\right)+\varepsilon\left(\pi^2-\frac 12\right)  u_t^2+  
Af^2 \right\} dx \nn
&\le& -c_3^2 d^2(u,u_t)+  \int_0^1Af^2 dx
\eea
where we have set
\beq
A:=\frac{\varepsilon}2+\frac 2{\varepsilon}, \qquad
c_3^2:=\frac{\omega_2}2\varepsilon,      \label{defs}
\eeq
and we have used (\ref{38}). In the sequel we shall
set also $p:=c_3^2/c_2^2$.

\subsection{Eventual boundedness and asymptotic stability}
\label{boundstab}

We assume that 
\beq
 A \int\limits_0^1 f^2 dx\,\leq g(t)c_1^2 d^2+
 \tilde g_1(t,d^2)+\tilde g_2(t,d^2),                      \label{hyp4'}
\eeq
where $f$ is the function appearing in $(\ref{31})$, and
$g(t),\tilde g_i(t,\eta)$ ($\ i=1,2$ and $ t\in J, \ \eta>0$)
denote suitable nonnegative continuous functions
fulfilling the following conditions:
\begin{itemize}

\item there exists a constant $\sigma>0$ such that
for any $t\ge t_0\ge 0$
\beq
\int\limits^t_{t_0}g(\tau)d\tau - p(t-t_0) \le \sigma;    \label{hyp1'}
\eeq

\item there exist
constants $\chi\in[0,1]$, $\kappa\in[0,1]$ and  $q\ge 0$ (with $q<p$ if 
$\chi=1$) and   $M>0$ such that 
\beq
\left|\frac{\int^t_0g(\tau)d\tau }{1+t^{\chi}}-q\right|< 
\frac M{1+t^{\kappa}};                           \label{hyp2'}
\eeq

\item for any $\eta>0$
\beq
\ba{l}
\lim\limits_{t\to+\infty}\tilde g_1(t,\eta)e^{\xi (t^{\chi}-t^{\kappa})}=0\\
\int\limits^{\infty}_0\tilde g_2(\tau,\eta)e^{\xi (\tau^{\chi}-
\tau^{\kappa})}d\tau 
=:\sigma_2(\eta)<+\infty,             
\ea                                                      \label{hyp3'}
\eeq
where $\xi$ is some positive constant if $\chi>\kappa$, 
while $\xi=0$ if  $\chi\le \kappa$.

\end{itemize}

Without loss of generality we can assume that 
$\tilde g_i(t,\eta)$
are non-decreasing in $\eta$; if originally this is not the case,
we just need to replace $\tilde g_i(t,\eta)$ by
$\max\limits_{0\le u\le \eta}\tilde g_i(t,u)$ to fulfill
this condition.

From (\ref{inter}), using (\ref{33}), (\ref{hyp4'}), (\ref{invemag}) 
we now find
\bea
\dot V(u,u_t) 
&\le & -c_3 d^2(u,u_t)+g(t)c_1^2d^2
+ \tilde{g}_1(t,d^2)+  \tilde{g}_2(t,d^2)\nn
&\le &
-p\, V+ g(t)V+ g_1(t,V)+g_2(t,V),
\eea
where we have set
\vskip.1cm
\beq 
 g_i(t,\eta):=\tilde{g}_i(t,\frac{\eta}{c_1^2})
\eeq
By the ``Comparison Principle''
(see e.g. \cite{yos}) $V$ is bounded from above 
\beq
V(t)\le y(t),            \label{mag1'}
\eeq
by the solution $y(t)$
of the Cauchy problem
\beq
\qquad\quad\dot y=-p\, y+ g(t)y+ g_1(t,y)+g_2(t,y), 
\qquad\qquad y(t_0)=y_0\equiv
V(t_0)\ge 0.
                                                    \label{eqconf'}
\eeq
We therefore study the latter, proving first a theorem
of eventual uniform boundedness.

\begin{lemma}
Assume $g(t),\tilde g_i(t,\eta)$ ($\ i=1,2$ and $ t\in J, \ \eta>0$)
are continuous nonnegative functions
fulfilling the conditions {\rm (\ref{hyp1'}-\ref{hyp3'})}. Then 
$\forall\tilde\alpha>0$ there exist $\tilde s(\tilde\alpha)\ge 0$, 
$\tilde\beta(\tilde\alpha)>0$
such that if $|y_0|\le\tilde\alpha$, $t_0\ge \tilde s(\tilde\alpha)$, the 
modulus of the solution
$y(t;t_0,y_0)$ of {\rm (\ref{eqconf'})} is bounded by $\tilde\beta$:
\beq
|y(t;t_0,y_0)|<\tilde\beta, \qquad\qquad t\ge t_0\ge \tilde s(\tilde\alpha);
\eeq
if in particular $y_0\in[0,\tilde\alpha]$, then
\beq                                                        \label{strbou}
0\le y(t;t_0,y_0)<\tilde\beta, \qquad\qquad t\ge t_0\ge \tilde s(\tilde\alpha).
\eeq
\label{lemma1'}
 \end{lemma}

\bp{}
Problem (\ref{eqconf'}) is equivalent to the integral equation
\bea
y(t) &=& y_0\,e^{- p(t-t_0)+\int\limits^t_{t_0}g(\tau)d\tau } \nn
&&\qquad +\, e^{- pt+\int\limits^t_0g(\tau)d\tau }
\int\limits_{t_0}^t\left[g_1\left(\tau,y(\tau)\right)+
g_2\left(\tau,y(\tau)\right)\right]e^{p\tau-
\int\limits_0^{\tau}g(z)dz}d\tau.               \label{solu'}
\eea
Take $\tilde\beta(\tilde\alpha):=\tilde\alpha(e^{\sigma}+\frac{e^{2M}}m+e^{2M})$, 
where
\beq                                                \label{defm}
m=\cases{ \frac p2 \quad\mbox{ if } \chi<1 \cr
          \frac {p-q}2\quad\mbox{ if } \chi=1.}
\eeq
Because
of (\ref{hyp1'}), if $|y_0|\le\tilde\alpha$, for any 
$t\ge t_0$ one finds
\beq
|y_0| e^{- p(t-t_0)+\int\limits^t_{t_0}g(\tau)d\tau }\le 
\tilde\alpha e^{\sigma}.                            \label{dis1'}
\eeq
Moreover, because of (\ref{hyp2'}), we obtain
\beq                                             \label{boa}
q(1+t^{\chi})-M\frac{1+t^{\chi}}{1+t^{\kappa}}< 
\int^t_0g(z)dz <q(1+t^{\chi})+M\frac{1+t^{\chi}}{1+t^{\kappa}}.
\eeq 
Let
$$
\begin{array}{l}
\vartheta:= \left\{
\begin{array}{ll}
0 \mbox{ if }\chi\le\kappa, \\[6pt]
\min\left\{1,\frac{\xi}{2M}\right\}
&\mbox{ if }1>\chi>\kappa \\[6pt]
\min\left\{1,\frac{p-q}{2M},\frac{\xi}{2M}
\right\} &\mbox{ if }1=\chi>\kappa,
\end{array}\right.\\[4pt]\\[4pt]
t_{\vartheta}:=\left\{
\begin{array}{ll}
0&\mbox{ if }\vartheta=0, \\[6pt]
\left(\frac{1-\vartheta}{\vartheta}\right)^{1/\kappa}&
\mbox{ if }\vartheta> 0;
\end{array}\right.
\end{array}
$$
considering separately the cases
$\chi\le \kappa$, $\chi> \kappa$
and recalling the definition of $\xi$, we find
$$
\frac{1+t^{\chi}}{1+t^{\kappa}}=1+\frac{t^{\chi}-t^{\kappa}}{1+t^{\kappa}}
\le 1+\vartheta (t^{\chi}-t^{\kappa}),
$$
for any $t\ge t_{\vartheta}$. Then from (\ref{boa})
\beq
\qquad q(1\!+\!t^{\chi})\!-\!M[1\!+\!\vartheta (t^{\chi}\!-\!t^{\kappa})]< 
\int^t_0g(z)dz <q(1\!+\!t^{\chi})+M[1\!+\!\vartheta (t^{\chi}\!-\!t^{\kappa})]  
                                              \label{double<'}
\eeq
for any $t\ge t_{\vartheta}$.
Consequently, for $i=1,2$ and $|y|\le \tilde\beta$
\beq \label{cicco'}
e^{- pt+\int\limits^t_0g(\tau)d\tau }
\int\limits_{t_0}^td\tau g_i(\tau,y)e^{p\tau-
\int\limits_0^{\tau}g(z)dz }d\tau  
\eeq
\bea
&& < e^{- pt+q(1+t^{\chi})+
M[1+\vartheta (t^{\chi}-t^{\kappa})]} \int\limits_{t_0}^tg_i(t,\tilde\beta)
e^{p\tau-q(1\!+\!\tau^{\chi})+M[1\!+\!\vartheta (\tau^{\chi}\!-\!
\tau^{\kappa})]}d\tau  \nn
&& =e^{qt^{\chi}\!+\!M\vartheta (t^{\chi}\!-\!t^{\kappa})
- pt}e^{2M}
\int\limits_{t_0}^tg_i(t,\tilde\beta)
e^{p\tau\!-\!q\tau^{\chi}\!+\!M\vartheta (\tau^{\chi}\!-\!
\tau^{\kappa})}d\tau,\nonumber
\eea
where we have used also the fact that $g_i(t,\eta)$ are
non-decreasing functions of $\eta$.

Now consider the function 
\beq                                            \label{defh}
h(\tau):=p\tau-q\tau^{\chi}-M\vartheta (\tau^{\chi}\!-\!\tau^{\kappa})
\eeq
and its derivative $h'(\tau)=p-q\chi\tau^{\chi-1}-M\vartheta(\chi
\tau^{\chi-1}-\kappa\tau^{\kappa-1})$.
We now show that, for any $\chi\in[0,1]$,
\beq
h'(\tau)\ge h'(\tilde t)=m \qquad\qquad \mbox{if }\tau\ge\tilde t:=
\left[\frac{\chi(2q+\xi)}p\right]^{\frac 1{1-\chi}}   \label{pippo'}
\eeq
with the $m$ defined in (\ref{defm}) [this implies that for 
$\tau\ge\tilde t$ the function $h(\tau)$ are increasing].
In fact, if $\vartheta> 0$, then it is either $0\le \kappa<\chi<1$, implying
$$
h'(\tau)>p-(q+M\vartheta)\chi\tau^{\chi-1}\ge\frac p2=m
$$
for any $\tau\ge\tilde t$,
or  $0\le \kappa<\chi=1$, implying (because of the inequality $p-q>0$
and the definition of $\vartheta$)
$$
h'(\tau)= p-q-M\vartheta +M\vartheta\kappa
\tau^{\kappa-1}>p-q-M\vartheta \ge\frac {p-q}2=m
$$
for any $\tau>0$, in particular for $\tau\ge\tilde t$.
If $\vartheta= 0$, then it is either $0\le \chi\le\kappa\le 1$
with $\chi<1$, implying
$$
h'(\tau)>p-q\chi\tau^{\chi-1}\ge\frac p2=m
$$
for any $\tau\ge\left[\frac{2q\chi}p\right]^{\frac 1{1-\chi}}\equiv\tilde t$,
or $\chi=\kappa= 1$, implying also $h'(\tau)=p-q>m$ (for any $\tau$),
as claimed.

On the other hand, because of (\ref{hyp3'}) there exist 
$s_1(\tilde\alpha),s_2(\tilde\alpha)\ge 0$ 
[recall that $\tilde\beta=\tilde\beta(\tilde\alpha)$] 
such that  
\beq
\ba{ll}
g_1(\tau,\tilde\beta)e^{\xi(\tau^{\chi}-\tau^{\kappa})} \le 
\tilde\alpha\qquad\qquad 
&\mbox{if}
\:\: \tau\ge t_0\ge s_1(\tilde\alpha), \\
\int\limits^{\infty}_{t_0}g_2(\tau,\tilde\beta)
e^{\xi(\tau^{\chi}-\tau^{\kappa})}d\tau\,  
\le \tilde\alpha\qquad\qquad &\mbox{if}
\:\: t_0\ge s_2(\tilde\alpha).
\ea                                             \label{chicco2}
\eeq
Hence, for
$t\ge t_0\ge \max\{\tilde t,t_{\vartheta}, s_1(\tilde\alpha)\}$ we find
that if $|y(\tau)|\le \tilde\beta$ for any $\tau\in[t_0,t[$ then
$$
e^{- pt+\int\limits^t_0g(\tau)d\tau }
\int\limits_{t_0}^tg_1\big(\tau,y(\tau)\big)e^{p\tau-
\int\limits_0^{\tau}g(z)dz }d\tau < e^{-h(t)+2M}
\int\limits_{t_0}^tg_1\big(\tau,\tilde\beta\big)
e^{h(\tau)+\xi(\tau^{\chi}-\tau^{\kappa})}d\tau 
$$
\bea
&&\qquad\qquad\le \:
e^{-h(t)+2M}\tilde\alpha\int\limits_{t_0}^t
\frac{h'(\tau)}m
e^{h(\tau)}d\tau = \:
\tilde\alpha\frac{e^{- h(t)+2M}}m (e^{h(t)}-e^{h(t_0)}) \nn
&&                                  \label{dis2'}
\qquad\qquad< \:\frac{e^{2M}}m \tilde\alpha     
\eea
where in the first line we have used (\ref{cicco'})
and the definition of $\vartheta$,
in the second (\ref{pippo'}) and (\ref{chicco2})$_1$.
Similarly, for $t\ge t_0\ge \max\{s_2(\tilde\alpha),t_{\vartheta},
\tilde t\}$ we find
that if $|y(\tau)|\le \tilde\beta$ for any $\tau\in[t_0,t[$ then
$$
e^{- pt+\int\limits^t_0g(\tau)d\tau }
\int\limits_{t_0}^tg_2\big(\tau,y(\tau))e^{p\tau-
\int\limits_0^{\tau}g(z)dz }d\tau < e^{-h(t)+2M}
\int\limits_{t_0}^tg_2\big(\tau,\tilde\beta)
e^{h(\tau)+\xi(\tau^{\chi}-\tau^{\kappa})}d\tau  
$$
\bea
&& \qquad\qquad< \: e^{-h(t)+h(t)+2M}
\int\limits^{\infty}_{t_0}\, 
g_2(\tau,\tilde\beta)e^{\xi(\tau^{\chi}-\tau^{\kappa})}d\tau  \nn
&& \qquad\qquad<\: \tilde\alpha e^{+2M},    \label{dis3'}
\eea
[in the first inequality we have used (\ref{cicco'}) and again 
the definition of $\vartheta$, in the 
second  the monotonicities of $h$ and $g_2$, in the
third  (\ref{chicco2})$_2$].
Summarizing, the inequalities (\ref{dis1'}), (\ref{dis2'}), (\ref{dis3'})
are fulfilled for
$t\ge t_0\ge \tilde s(\tilde\alpha):=\max\{\tilde t,t_{\vartheta},
s_1(\tilde\alpha),s_2(\tilde\alpha), \}$.

Now let us suppose {\it per absurdum} that there exists
$t_1> t_0\ge \tilde s(\tilde\alpha)$ such that 
\bea
&&|y(t;t_0,y_0)|<\tilde\beta \qquad\qquad\mbox{for } t_0\le t<t_1 
                                                            \label{ggg0'}\\
&&|y(t_1;t_0,y_0)|=\tilde\beta.                               \label{ggg'}
\eea
Because of (\ref{ggg0'}) the inequalities (\ref{dis2'}), 
(\ref{dis3'}) are fulfilled; together with Eq. (\ref{solu'}),
(\ref{dis1'}) for $t=t_1$ they imply
\[
|y(t_1;t_0,y_0)|<\tilde\beta,
\]
against the assumption (\ref{ggg'}). Finally, from (\ref{solu'}) and the
nonnegativity of the functions $g_i$ we find that $0\le y_0<\tilde\alpha$
implies $y(t)>0$ for any $t$, whence (\ref{strbou}). \ep

As a result of the previous lemma, for any
$\tilde\alpha>0$ the solution $y(t)$ of the
Cauchy problem (\ref{eqconf'}) remains eventually
uniformly bounded by $\tilde\beta(\tilde\alpha)$
if $0\le y_0\le\tilde\alpha$. By  (\ref{mag1'}) and (\ref{invemag}), 
the same applies with $V(t)$ and $d^2(u,u_t)$.

By the monotonicity of $g_i(t,\eta)$ in $\eta$ and the comparison
principle we find that $y(t)$ is also bounded 
\vskip0.1cm
\beq
y(t)\le z(t) \qquad\qquad t\ge t_0                   \label{mag2'}       
\eeq
by the solution $z(t)$
of the Cauchy problem 
\beq
\dot z=-p\, z+ g(t)z+ g_1(t,\tilde\beta)+g_2(t,\tilde\beta), \qquad\qquad 
z(t_0)=z_0                                     \label{eqconf2'}
\eeq
[which differs from (\ref{eqconf'}) in that the second
argument of $g_i$ is now a fixed constant $\tilde\beta>0$], provided that
$z_0=y_0$, and $t_0\ge \tilde s(\tilde\alpha)$. 

We therefore study the Cauchy problem (\ref{eqconf2'}), 
keeping in mind that for our final purposes we will choose
$\tilde\beta=\tilde\beta(\tilde\alpha)$, 
$t_0=t_0(\tilde\alpha)\ge \tilde s(\tilde\alpha)$.

\begin{lemma}
Assume $g(t),\tilde g_i(t,\eta)$ ($\ i=1,2$ and $ t\in J, \ \eta>0$)
are continuous functions
fulfilling the conditions {\rm (\ref{hyp1'}-\ref{hyp3'})}. Then for any
$\tilde\rho>0$, $t_0>0$, $\tilde\alpha>0$ 
there exists $\hat T(\tilde\rho,\tilde\alpha,\tilde\beta,t_0)>0$
such that for $|z_0|\le\tilde\alpha$ $ \in[0,\tilde\alpha]$ the solution 
$z(t;t_0,z_0)$ of {\rm (\ref{eqconf2'})} is bounded as follows:
\beq
|z(t;t_0,z_0)|<\tilde\rho, \qquad\qquad \mbox{if } t\ge t_0+\hat T.
\eeq
If in particular $ z_0\in[0,\tilde\alpha[$, then
\beq
0\le z(t;t_0,z_0)<\tilde\rho, \qquad\qquad \mbox{if } 
t\ge t_0+\hat T. \label{zaza'}
\eeq
\label{lemma2'}
 \end{lemma}

\bp{}
The solution $z(t)=z(t;t_0,z_0)$ is of the form
\bea
z(t) &=&  
z_0\,e^{- p(t-t_0)+\int\limits^t_{t_0} g(\tau)d\tau} \nn
&&+\, e^{- pt+\int\limits^t_0g(\tau)d\tau }
\int\limits_{t_0}^t\left[g_1(\tau,\tilde\beta)+
g_2(\tau,\tilde\beta)\right]e^{p\tau-\int\limits_0^{\tau}g(\lambda) 
d\lambda }d\tau.                                     \label{solu2'}
\eea
We now consider each of the three terms at the rhs of (\ref{solu2'})
separately. 

By eq.  (\ref{double<'}) for $t\ge t_{\vartheta}$
$$ 
-p(t\!-\!t_0)+\int\limits^t_{t_0}g(\tau)d\tau \le
-(t\!-\!t_0) \left[p-q\frac{1+t^{\chi}}{t-t_0}-M
\frac{1+\vartheta (t^{\chi}\!-\!t^{\kappa})}{t\!-\!t_0}\right];
$$
the right hand-side is negatively divergent for $t-t_0\to+\infty$,
and so will be the left hand-side;
this implies that there exists a 
$T_0(\tilde\rho,\tilde\alpha,t_0)\ge 0$ such that 
\beq
|z_0|\,e^{- p(t-t_0)+\int\limits^t_{t_0} g(\tau)d\tau}\, <\,\frac{\tilde\rho}3,
\qquad\qquad t\ge t_0+ T_0,\: z_0\in[-\tilde\alpha,\tilde\alpha].    
\label{magg1'}
\eeq

As for the second term, 
given $\tilde\beta>0,\tilde\rho>0$, because of (\ref{hyp3'})$_1$ there
exist 
$T_1(\tilde\beta,\tilde\rho)\ge 
\max\{\tilde t,t_{\vartheta}\}$ 
and $\sigma_1(\tilde\beta)$ such that
\beq
\ba{ll}
g_1(\tau,\tilde\beta) \le \sigma_1\qquad\qquad &\mbox{if}
\:\: \tau\ge 0, \\[8pt]
g_1(\tau,\tilde\beta)e^{\xi(\tau^{\chi}-\tau^{\kappa})} \le \frac 16\,
m\tilde\rho \,e^{-2M}
\qquad\qquad &\mbox{if}\:\: \tau\ge T_1 
\ea                                             \label{chiccoo'}
\eeq
[$\tilde t,m$ have been defined resp. in (\ref{pippo'}),  (\ref{defm})].
Since the function $h(t)$ defined in (\ref{defh}) is increasing
as the first power of $t$ for $t\ge \tilde t$, there exists 
$T_2(\tilde\beta,\tilde\rho)\ge T_1$ such that for $t\ge T_2$
\beq
\frac{\sigma_1}p\,e^{-h(t)+M+q+pT_1}
<\frac{\tilde\rho}6.                                 \label{mmmm}
\eeq
Therefore, for $t\ge\ T_2$,
\bea
&&e^{- pt+\int\limits^t_0g(\tau)d\tau }
\int\limits_{t_0}^tg_1(\tau,\tilde\beta)e^{p\tau-
\int\limits_0^{\tau}g(\lambda)d\lambda }d\tau \nn 
&&\qquad\qquad < \:\: e^{- pt+q(1+t^{\chi})+M[1+
\vartheta(t^{\chi}-t^{\kappa})]}
\int\limits_0^tg_1(\tau,\tilde\beta)e^{p\tau-
\int\limits_0^{\tau}g(\lambda)d\lambda }d\tau \nn 
&&\qquad\qquad<\:\:e^{-h(t)\!+\!M\!+\!q\!}
\int\limits_0^{T_1}g_1(\tau,\tilde\beta)
e^{p\tau}d\tau 
+ e^{-h(t)\!+\!2M\!}
\int\limits^t_{T_1}g_1(\tau,\tilde\beta)
e^{h(\tau)\!+\!\xi(\tau^{\chi}\!-\!\tau^{\kappa})}d\tau\nn
&& \qquad\qquad< \:\:e^{-h(t)+M+q}\sigma_1
\int\limits_0^{T_1} e^{p\tau}d\tau +\:\: e^{-h(t)+2M }
e^{-2M}\frac{m\tilde\rho}6 
\int\limits^t_{T_1}\frac{h'(\tau)}m 
e^{h(\tau)}d\tau  \nn
&&\qquad\qquad<\:\:e^{-h(t)+M+q}\sigma_1\frac{e^{pT_1}}p
 + \frac{\tilde\rho}6 e^{-h(t)}(e^{h(t)}-e^{h(T_1)}) \nn
&&                                              \label{dis2z'}
\qquad\qquad< \:\: \frac{\tilde\rho}6(1+1)=\frac{\tilde\rho}3,       
\eea
where in the first and in the second inequality we have used
(\ref{double<'}), the nonnegativity of $g_1$, the fact that
$\xi(\tau^{\chi}-\tau^{\kappa})\ge 0$ and
the definition of $h(t)$, in the third (\ref{chiccoo'})
and (\ref{pippo'}), 
in the fourth we have performed integration 
over $\tau$, and in the last we have used (\ref{mmmm}).

As for the third term at the rhs of (\ref{solu2'}), 
from (\ref{hyp3'})$_2$ it follows that there exists
$T_3(\tilde\beta,\tilde\rho)\ge 
\max\{\tilde t,t_{\vartheta}\}$ 
such that for 
$t\ge T_3$
\beq
e^{2M}\int\limits^t_{T_3}g_2(\tau,\tilde\beta)
e^{\xi (\tau^{\chi}-\tau^{\kappa})}d\tau 
<\frac{\tilde\rho}6                                \label{chiccoo"}
\eeq
and on the other hand that
\beq
\int\limits^{T_3}_0g_2(\tau,\tilde\beta)e^{p\tau}
d\tau <\sigma_2,                          \label{chiccoo"'}
\eeq
where $\sigma_2$ has been defined in (\ref{hyp3'}).
Moreover, there exists 
$T_4(\tilde\beta,\tilde\rho)\ge T_3$ 
such that for $t\ge\tilde t+ T_4$
\beq
\sigma_2e^{-h(t)+q+M}<\frac{\tilde\rho}6.     \label{mmmm'}
\eeq
Therefore for $t\ge T_4$
$$
e^{- pt+\int\limits^t_0 g(\tau)d\tau}
\int\limits_{t_0}^tg_2(\tau,\tilde\beta)e^{p\tau-
\int\limits_0^{\tau}g(z)dz }d\tau
$$
\bea
&&\qquad \qquad<\:\: e^{- pt+q(1+t^{\chi})+M[\vartheta(t^{\chi}-t^{\kappa})+1]}
\int\limits_0^tg_2(\tau,\tilde\beta)e^{p\tau-
\int\limits_0^{\tau}g(z)dz }d\tau \nn
&& \qquad\qquad < 
\:\:e^{- h(t)+q+M}
\int\limits_0^{T_3}g_2(\tau,\tilde\beta)
e^{p\tau }d\tau
+ e^{- h(t)+2M}\int\limits_{T_3}^tg_2(\tau,\tilde\beta) 
e^{\xi (\tau^{\chi}-\tau^{\kappa})} e^{h(\tau)}d\tau \nn
&&\qquad\qquad < 
\:\:e^{- h(t)+q+M}\sigma_2
+ e^{- h(t)+2M+h(t)}\int\limits_{T_3}^tg_2(\tau,\tilde\beta) 
e^{\xi (\tau^{\chi}-\tau^{\kappa})}d\tau \nn
&&\label{dis3z'}\qquad\qquad< \:\: \frac{\tilde\rho}6 +
\frac{\tilde\rho}6=\frac{\tilde\rho}3, 
\eea
where we have used  the
nonnegativity of $g_2$ and (\ref{double<'})  in the first inequality, 
again (\ref{double<'}), the fact that
$\xi(\tau^{\chi}-\tau^{\kappa})\ge 0$ and the
nonnegativity of $g$ in the second, (\ref{chiccoo"'}) and 
the monotonicity of $h(\tau)$ for $\tau\ge T_3$
in the third,  (\ref{chiccoo"}) and (\ref{mmmm'}) in the last one.

Let 
$\hat T(\tilde\rho,\tilde\beta,t_0):=\max\{T_0,T_2,T_4\}$. 
Collecting
the results (\ref{magg1'}), (\ref{dis2z'}), (\ref{dis3z'}) we find that
the solution $z(t)$ of (\ref{eqconf2'}) fulfills the condition
\[
z(t,t_0,z_0)<\frac{\tilde\rho}3[1+1+1]=\tilde\rho, \qquad\qquad t\ge 
t_0+\hat T.
\]
\ep

{\bf Remark 1.} This lemma is a generalization of Lemma 24.3
in \cite{yos}, based in turn on an argument due to Hale \cite{Hale}.

\bigskip
{\bf Remark 2.}
If $\chi\le \kappa$ then in the previous proof
$T_0$, and therefore $\hat T$, becomes independent of $t_0$. In fact, 
$\vartheta=0$ and from (\ref{double<'}) we find
$$
\int^t_{t_0}g(z)dz =\int^t_0g(z)dz -\int^{t_0}_0g(z)dz
 <q(t^{\chi}-t_0^{\chi})+2M.
$$
By Lagrange's theorem there exists a $\tau\in]t_0,t[$ such that
$t^{\chi}-t_0^{\chi}=\frac{\chi}{\tau^{1-\chi}}(t-t_0)$.
Since $t_0\ge \tilde t\equiv(2q\chi/p)^{1/(1-\chi)}$
we find
$t^{\chi}-t_0^{\chi}<\frac{p}{2q}(t-t_0)$
$$
-p(t-t_0)+\int^t_{t_0}g(z)dz 
 <-p(t-t_0)\left[1-\frac 12\right]+2M=-\frac p2(t-t_0)+2M.
$$
This implies that the left-hand is negatively divergent for
$t-t_0\to +\infty$ uniformly in $t_0$, as anticipated.
The argument is not applicable in the case $\chi>\kappa$.
\bigskip

We are now in the conditions to prove the following

\begin{theorem}
Assume that the function $f$ of {\rm (\ref{31})} is bounded as
in {\rm (\ref{hyp4'})}, where
$g(t),\tilde g_i(t,\eta)$ ($\ i=1,2$ and $ t\in J, \ \eta>0$)
are continuous functions
fulfilling the conditions {\rm (\ref{hyp1'}-\ref{hyp3'})}. Then the
solutions of the problem {\rm  (\ref{31}), (\ref{32})} are eventually
uniformly bounded. Moreover, the origin $O$  is eventually
quasi-uniform-asymptotically stable in the large with respect
to the metric $d$. 
\end{theorem}

\bp{}
Set $\tilde\alpha:=\alpha^2 c_2^2$, and apply lemma \ref{lemma1'}.
Under the assumption $d(u_0,u_1)\le \alpha$, by 
(\ref{invemag}) we find $y_0=V(t_0)\le\tilde\alpha$, by
(\ref{mag1'}) and the application of the lemma 
we find that $y(t)$ (and therefore $V(t)$)
is bounded by $\tilde\beta(\alpha^2 c_2^2)$, and again by
(\ref{invemag}) we find 
$d(t)\le \beta(\alpha):=\sqrt{\tilde\beta(\alpha^2 c_2^2)/c^2_1}$
for $t\ge s(\alpha):=\tilde s(\alpha^2 c_2^2)$,
as claimed. Moreover, we can now apply
the comparison principle (\ref{mag2'}-\ref{eqconf2'}) and
lemma \ref{lemma2'}: chosen $\rho>0$, we set
$\tilde\rho:=c_1^2\rho^2$.
As a consequence of (\ref{mag2'}), (\ref{zaza'}), 
(\ref{invemag}) we thus find that for $t_0\ge s(\alpha)$
and 
$t\ge \hat T(c_1^2\rho,t_0(\alpha),c_2^2\alpha^2)\equiv T(\rho,\alpha)$
$$
d^2(t)\le \frac{V(t)}{c_1^2}\le\frac{y(t)}{c_1^2}
\le\frac{z(t,\tilde\beta(\alpha))}{c_1^2}<\frac{\tilde\rho}{c_1^2}
=\rho^2.
$$
\ep

{\bf Remark 3.}
This theorem is a generalization of Thm 3.1. in reference
\cite{DacDan98}: the claims are the same, but
the hypotheses on the function $f$
are weakened. First, (\ref{hyp4'}) is an upper bound
condition only on the mean square value of $f^2$, rather than
on its supremum (as in \cite{DacDan98}). Second, this upper bound
may depend on $t$ in a more general way than in
that reference. The hypotheses
(\ref{hyp1'}), (\ref{hyp2'}), (\ref{hyp3'}) considered here 
are fulfilled by the ones considered there with $g(t)\equiv$const
and $\chi=\kappa=1$. The former, but not the latter,
are satisfied e.g. by the following family of

\bigskip

{\bf Examples.} Let
$f=b(t)\sin\varphi$, with a function $b(t)$ such that the integral
$\int_0^t b^2(\tau)d\tau$ grows as some power $t^{\chi}$, where
$\chi\le 1$, and in the case $\chi=1$ is smaller than $pt$ for
sufficiently large $t$; then we can set $\hat g(t,\eta)\equiv b^2(t)$.
For instance we could take $b^2$ a continuous function that 
vanishes everywhere
except in intervals centered, say, at equally spaced points, where 
it takes maxima increasing with some power law $\sim t^{\beta}$,
but keeps the integral bounded, e.g.
\beq
\qquad\quad b^2(t)=b_0^2\:\left\{
\begin{array}{ll}
4n^{\alpha+\beta}(t-n+\frac 1{2n^{\alpha}})\qquad & \qquad\mbox{if} 
\qquad t\in[n-\frac 1{2n^{\alpha}},n],\cr
4n^{\beta}- 4n^{\alpha+\beta}(t-n) \qquad &\qquad\mbox{if} \qquad 
t\in ]n,n+\frac 1{2n^{\alpha}}],\cr      0\qquad &\qquad\mbox{otherwise,}
\end{array}\right.
\eeq
with $b_0^2<p$, $\alpha\ge 1$, $\beta\in]\alpha-1,\alpha]$
and $n\in\b{N}$. (The case $\alpha=\beta=1$ has already been 
considered in  \cite{DanFio00}). 

The graph
of $(b(t)/b_0)^2$ consists of a sequence of
isosceles triangles enumerated by $n$,
having bases of lenght $1/n^{\alpha}$ and upper vertices with
coordinates $(x,y)=(n,2n^{\beta})$ (see the figure). Their
areas are $A_n=1/n^{\gamma}$, where $\gamma:=\alpha-\beta\in[0,1[$.

If $0\le t-t_0< 2$ then we immediately find
\beq
\int\limits^t_{t_0} g(\tau)d\tau\le b_0^2\, 2.       \label{facile}
\eeq
If on the contrary if $t-t_0\ge 2$ then there exist integers $m,n$
with $0\le m\le n-2$ and $t> t_0\ge 0$ such that
$t\in]n-1/2,n+1/2]$ and
$t_0\in]m-1/2,m+1/2]$. Then we find
$$
\int\limits^{n-1/2}_{m+1/2} g(\tau)d\tau\le
\int\limits^t_{t_0} g(\tau)d\tau\le
\int\limits^{n+1/2}_{m-1/2} g(\tau)d\tau,
$$
namely
\beq
\sum\limits^{n-1}_{k=m+1}\frac{b_0^2}{k^{\gamma}}=
b_0^2\sum\limits^{n-1}_{k=m+1}A_k \le
\int\limits^t_{t_0} g(\tau)d\tau\le b_0^2\sum\limits^n_{k=m\atop k\ge 1}A_k
=\sum\limits^n_{k=m\atop k\ge 1}\frac {b_0^2}{k^{\gamma}}.     \label{dudu}
\eeq

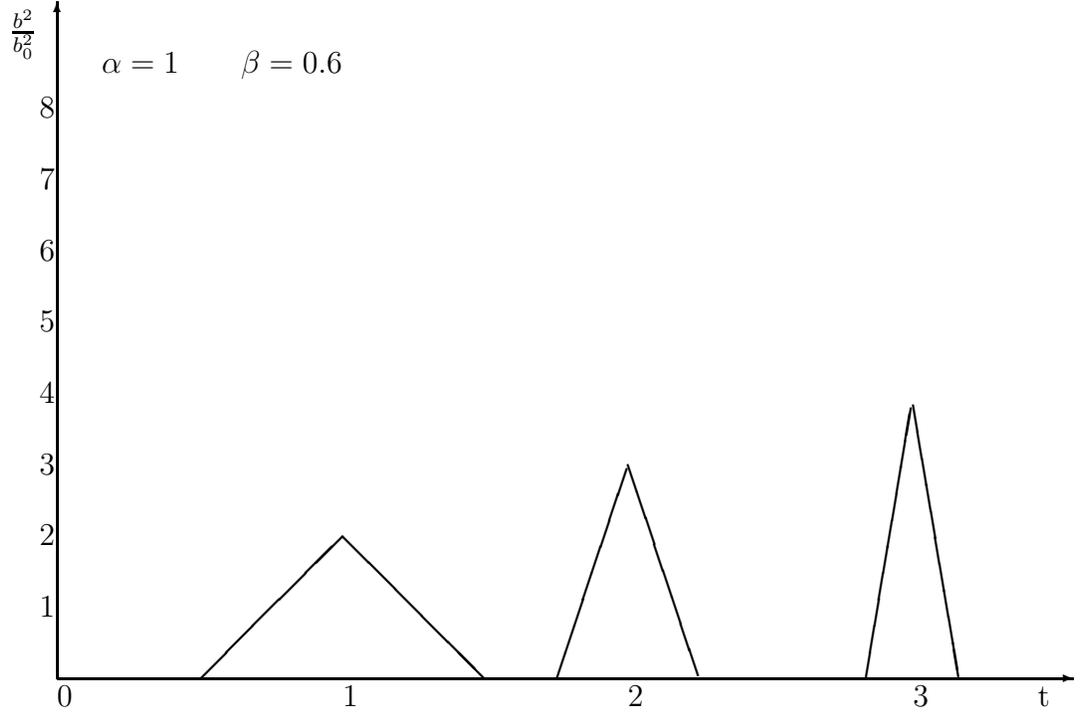
\setlength{\unitlength}{2947sp}%
\begingroup\makeatletter\ifx\SetFigFont\undefined%
\gdef\SetFigFont#1#2#3#4#5{%
  \reset@font\fontsize{#1}{#2pt}%
  \fontfamily{#3}\fontseries{#4}\fontshape{#5}%
  \selectfont}%
\fi\endgroup%
\begin{picture}(9012,5937)(151,-5386)
\thinlines
\put(451,-5161){\vector( 0, 1){5700}}
\put(451,-5161){\vector( 1, 0){8550}}
\thicklines
\put(451,-5161){\line( 1, 0){1200}}
\put(1651,-5161){\line( 1, 1){1200}}
\put(2851,-3961){\line( 1,-1){1200}}
\put(4051,-5161){\line( 1, 0){600}}
\put(4651,-5161){\line( 1, 3){590}}
\put(5251,-3361){\line( 1,-3){590}}
\put(5851,-5161){\line( 1, 0){1400}}
\put(7251,-5161){\line( 1, 6){380}}
\put(7651,-2861){\line( 1,-6){380}}
\put(8051,-5161){\line( 1, 0){575}}
\put(451,-5386){\makebox(0,0)[lb]{\smash{\SetFigFont{12}{14.4}{\rmdefault}{\mddefault}{\updefault}
0
}}}
\put(301,-4636){\makebox(0,0)[lb]{\smash{\SetFigFont{12}{14.4}{\rmdefault}{\mddefault}{\updefault}
1
}}}
\put(301,-4036){\makebox(0,0)[lb]{\smash{\SetFigFont{12}{14.4}{\rmdefault}{\mddefault}{\updefault}
2
}}}
\put(301,-3436){\makebox(0,0)[lb]{\smash{\SetFigFont{12}{14.4}{\rmdefault}{\mddefault}{\updefault}
3
}}}
\put(301,-2836){\makebox(0,0)[lb]{\smash{\SetFigFont{12}{14.4}{\rmdefault}{\mddefault}{\updefault}
4
}}}
\put(301,-2236){\makebox(0,0)[lb]{\smash{\SetFigFont{12}{14.4}{\rmdefault}{\mddefault}{\updefault}
5
}}}
\put(301,-1636){\makebox(0,0)[lb]{\smash{\SetFigFont{12}{14.4}{\rmdefault}{\mddefault}{\updefault}
6
}}}
\put(301,-1036){\makebox(0,0)[lb]{\smash{\SetFigFont{12}{14.4}{\rmdefault}{\mddefault}{\updefault}
7
}}}
\put(301,-436){\makebox(0,0)[lb]{\smash{\SetFigFont{12}{14.4}{\rmdefault}{\mddefault}{\updefault}
8
}}}
\put(826,-61){\makebox(0,0)[lb]{\smash{\SetFigFont{12}{14.4}{\rmdefault}{\mddefault}{\updefault}
$\alpha= 1\qquad    \beta=0.6$
}}}
\put(051,214){\makebox(0,0)[lb]{\smash{\SetFigFont{12}{14.4}{\rmdefault}{\mddefault}{\updefault}
$\frac{b^2}{b_0^2}$
}}}
\put(2851,-5386){\makebox(0,0)[lb]{\smash{\SetFigFont{12}{14.4}{\rmdefault}{\mddefault}{\updefault}
1
}}}
\put(5251,-5386){\makebox(0,0)[lb]{\smash{\SetFigFont{12}{14.4}{\rmdefault}{\mddefault}{\updefault}
2
}}}
\put(7651,-5386){\makebox(0,0)[lb]{\smash{\SetFigFont{12}{14.4}{\rmdefault}{\mddefault}{\updefault}
3
}}}
\put(8701,-5386){\makebox(0,0)[lb]{\smash{\SetFigFont{12}{14.4}{\rmdefault}{\mddefault}{\updefault}
t
}}}
\end{picture}

\bigskip

Consider the function $e(y):=y^{1-\gamma}$, $\gamma\in[0,1[$.
Applying Lagrange's theorem we find that 
for any $h\in\b{N}$ there exists a $\xi_h\in]h,h+1[$
such that
$$
(h+1)^{1-\gamma}-h^{1-\gamma}=(1-\gamma)\frac1{\xi_h^{\gamma}},
$$
whence, taking $h=k$ and $h=k-1$ respectively,
\bea
&&(k+1)^{1-\gamma}-k^{1-\gamma}<(1-\gamma)\frac1{k^{\gamma}},\nn
&&k^{1-\gamma}-(k-1)^{1-\gamma}>(1-\gamma)\frac1{k^{\gamma}};
\nonumber
\eea
therefore
\beq
\frac1{1\!-\!\gamma}[(k\!+\!1)^{1-\gamma}-k^{1-\gamma}]<
\frac 1{k^{\gamma}}<
\frac1{1\!-\!\gamma}[k^{1-\gamma}-(k\!-\!1)^{1-\gamma}]. \label{dada}
\eeq
From (\ref{dudu}), (\ref{dada}) we find
\beq
\qquad \frac{b_0^2
[n^{1-\gamma}-(m\!+\!1)^{1-\gamma}]}{1\!-\!\gamma}<
\int\limits^t_{t_0} g(\tau)d\tau<
\frac{b_0^2[n^{1-\gamma}-
(m\!-\!1)^{1-\gamma}(1\!-\!\delta_0^m)]}{1\!-\!\gamma}.    \label{bound}
\eeq
where $\delta_0^m$ denotes a Kronecker $\delta$. Hence,
\beq
\int\limits^t_{t_0} g(\tau)d\tau= 
\frac{b_0^2}{1\!-\!\gamma}[n^{1-\gamma}-(m\!+\!1)^{1-\gamma}]+
L_{m,n}(t)                                 \label{remainder}
\eeq
where the remainder $L_{m,n}(t)$ is bound by the difference $d_m$
of the rhs and lhs of (\ref{bound}),
$$
0< L_{m,n}(t)< d_m:=\frac{b_0^2}{1\!-\!\gamma}[
(m\!+\!1)^{1-\gamma}-(m\!-\!1)^{1-\gamma}(1\!-\!\delta_0^m)]. 
$$
The expression in square bracket equals 1 for $m=0$ and $2^{1-\gamma}$
for $m=1$.
It is immediate to check that the function 
$\tilde e(y):=(y\!+\!1)^{1-\gamma}-(y\!-\!1)^{1-\gamma}$ is decreasing
for $y\ge 1$ and therefore takes its maximum in $y=1$. We 
therefore derive the bound 
\beq
0< L_{m,n}(t)<d_m\le \frac{b_0^2\,\tilde e(1)}{1\!-\!\gamma}=
\frac{b_0^2\,2^{1-\gamma}}{1\!-\!\gamma}.
                                              \label{bound'}
\eeq
Moreover, since $t>n\!-\!1$, $t_0<m\!+\!1$ and $g$ is nonnegative, 
from (\ref{remainder}) we find
$$
\int\limits^t_{t_0} g(\tau)d\tau<
\frac{b_0^2}{1\!-\!\gamma}[(t\!+\!1)^{1-\gamma}-t_0^{1-\gamma}]+
L_{m,n}(t). 
$$
If $t_0\ge 1$,
applying again Lagrange's theorem to the function 
$e(t)=t^{1-\gamma}$ we find
$$
\frac{b_0^2}{1\!-\!\gamma}[(t\!+\!1)^{1-\gamma}-t_0^{1-\gamma}]
=b_0^2\frac{t-t_0+1}{\bar t{}^{\gamma}}< b_0^2(t-t_0+1)
$$
with a suitable $\bar t\in]t_0,t+1[$, and therefore
\beq
\int\limits^t_{t_0} g(\tau)d\tau-b_0^2(t-t_0)<b_0^2\left(1+
\frac{2^{1-\gamma}}{1\!-\!\gamma}\right)           \label{allegro}
\eeq
If $0\le t_0<1$, 
$$
\int\limits^t_{t_0} g(\tau)d\tau-b_0^2(t-t_0)\le
\int\limits^1_0 g(\tau)d\tau-b_0^2(1-t_0)+
\int\limits^t_1 g(\tau)d\tau-b_0^2(t-1)
<b_0^2\left(2+
\frac{2^{1-\gamma}}{1\!-\!\gamma}\right)=:\sigma,
$$
where we have used (\ref{allegro}) with $t_0=1$
and $\int\limits^1_0g(\tau)d\tau\le b_0^2$,
showing [together with (\ref{allegro}) itself and
(\ref{facile})]
that $g$ fulfills condition (\ref{hyp1'})
in any case. 

On the other hand,
choosing $t_0=0$ (and therefore $m=0$) 
in (\ref{remainder}), dividing
by $1\!+\!t^{1-\gamma}$ and subtracting $b_0^2/(1\!-\!\gamma)$ we find 
$$
\frac{\int\limits^t_0 g(\tau)d\tau}{1+t^{1-\gamma}}
-\frac{b_0^2}{1\!-\!\gamma}= 
\frac{b_0^2}{1\!-\!\gamma}\left[\frac{n^{1-\gamma}
-(1\!+\!t^{1-\gamma})-1}{1+t^{1-\gamma}}
\right]+\frac{L_{0,n}(t)}{1+t^{1-\gamma}}
$$
But it is $n\!-\!1<t<n\!+\!1$, what  implies 
$$ 
1-2^{1-\gamma}\le n^{1-\gamma}\!-(n\!+\!1)^{1-\gamma}
 <n^{1-\gamma}\!-\!t^{1-\gamma}
<(t\!+\!1)^{1-\gamma}\!-\!t^{1-\gamma}< 1
$$
(in fact the  function $\hat e(y):=(y\!+\!1)^{1-\gamma}
\!-\!y^{1-\gamma}$ is decreasing and therefore has
maximum at the lower extremum of the interval in
which we define it); hence, using
also (\ref{bound'}), we find
$$
-\frac{b_0^2}{1\!-\!\gamma}\left[\frac{2^{1-\gamma}+1}{1+t^{1-\gamma}}\right]<
\frac{\int\limits^t_0 g(\tau)d\tau}{1+t^{1-\gamma}}
-\frac{b_0^2}{1\!-\!\gamma}<\frac{b_0^2}{1\!-\!\gamma}\left[\frac{2^{1-\gamma}
-1}{1+t^{1-\gamma}}\right]<\frac{b_0^2}{1\!-\!\gamma}\left[\frac{2^{1-\gamma}
+1}{1+t^{1-\gamma}}\right].
$$
We have proved these inequalities under the current assumption $t\ge 2$,
showing that in this domain also condition (\ref{hyp2'}),
with $q=b_0^2/(1\!-\!\gamma)$, $\chi=\kappa=1-\gamma$ and
$M=b_0^2(2^{1-\gamma}\!+\!1)/(1\!-\!\gamma)$, is satisfied.
For $0\le t\le 2$ the left-hand side of (\ref{hyp2'}) is certainly bounded
by $b_0^2\,3/[2(1-\gamma)]$, therefore it is sufficient to choose e.g.
$M=b_0^2\,9/[2(1-\gamma)]$ to fulfill  (\ref{hyp2'}) for any $t\ge 0$.

\sect{Exponential-asymptotic stability for special $f$'s via a family
of Liapunov functionals}
\label{expstab}

In this section we specialize the function $f$ of (\ref{31}) as
$f=F(u)\!-\!a(x\!,\!t\!,\!u\!,\!u_x\!,\!u_t\!,\!u_{xx})u_t$, 
where $F\in C(\b{R})$ 
and $a\in C(]0,1[\times J \times \b{R}^4)$, 
and examine the particular problem
  \beq                               \label{320}
 \left\{
   \begin{array}{l}
    Lu=F(u)-a(x,t,u,u_x,u_t,u_{xx})u_t, \qquad x\in ]0,1[,\:\: t>t_0
                                                 \vspace{3mm}\\
    u(0,t)=0, ~~u(1,t)=0, \qquad \qquad t>t_0
   \end{array}
 \right.
 \eeq
with initial and consistency conditions  (\ref{32}-\ref{32bis}).
We shall use the one-parameter family of modified Liapunov functionals
\bea                                               \label{321}
W_{\gamma}(\varphi,\psi) &=&
\frac{1}{2}\int_0^1 \left\{(\varepsilon \varphi_{xx}-\psi)^2+
\gamma\psi^2+(1\!+\!\gamma)\varphi_x^2\right\}dx\\
&& -(1\!+\!\gamma)\int_0^1
\left(\int_0^{\varphi(x)}F(z)dz\right)dx \nonumber
\eea
where $\gamma>1/2$ for the moment is an unspecified
parameter.

\begin{theorem}
Under the following assumptions
\begin{itemize}
\item $F(u)\in C^1(\b{R}), \ F(0)=0$, and moreover
 there exists a positive constant $K$ such that
\beq 
F_u\leq K<3\pi^2/4                                    \label{HYP1}
\eeq
\item the function $a$ satisfies
\beq
\nu:= \varepsilon\pi^2+ \inf a > 0;      \label{324}
\eeq
\item there exist $\tau\in[0,2[$ and constants $A>0,A'\ge 0$
such that
\beq 
a(x,t,\varphi,\varphi_x,\varphi_{xx},\psi)\le 
A\left[d(\varphi,\psi)\right]^{\tau}+A', 
\label{HYP3}
\eeq
\end{itemize}
the null solution of the problem
(\ref{320}) is exponential-asymptotically stable in the large.
\end{theorem}
As anticipated in the introduction,
this should be compared with Thm 3.3. in the main reference, 
\cite{DacDan98}: by replacing the requirement
that $\sup a<\infty$ and adding the assumption
(\ref{HYP3}) we are still able to prove the  
exponential-asymptotic stability in the large of the null
solution.
The trick is to associate to each neighbourhood of the origin 
with radius $\sigma$
(the `error') a Liapunov functional
(\ref{321}) with parameter $\gamma$ adapted to $\sigma$,  instead
of fixing $\gamma$ once and for all.

\bp{}  We start by improving or recalling some inequalities proved
in \cite{DacDan98}. From (\ref{HYP1}) we find
\beq \label{zzz}
\int_0^{\varphi}F(z)dz=\int_0^{\varphi}dz\int_0^z F_s(s)ds\le 
K\int_0^{\varphi}dz\int_0^zds=K\varphi^2/2.
\eeq
Employing this inequality and the estimate (\ref{38})
we find
\bea 
W_{\gamma}(\varphi,\psi) 
= \frac{1}{2}\int_0^1  \Big\{(\varepsilon \varphi_{xx}-2\psi)^2/4+
(\varepsilon \varphi_{xx}-\psi)^2/2+(\gamma-1/2)\psi^2\nn
\qquad \qquad + (1+\gamma)\varphi_x^2  +\varepsilon^2 \varphi_{xx}^2/4 
-2(1\!+\!\gamma)\int_0^{\varphi}F(z)dz \Big \}dx \nonumber
\eea
\beq  \label{lal}
\qquad \ge \frac{1}{2}\int_0^1  \left[\left(\gamma\!-\!\frac 12
\right)\psi^2+(1+\gamma)
\varphi_x^2+\frac{\varepsilon^2}4 \varphi_{xx}^2
-2(1\!+\!\gamma)\int_0^{\varphi}F(z)dz  \right]dx 
\eeq
$$
\ge \frac{1}{2}\int_0^1\left[\left(\gamma\!-\!
\frac 12\right)\psi^2+(1+\gamma)\pi^2
\varphi^2+\frac{\varepsilon^2}4\omega_3 (\varphi_{xx}^2 +\varphi_x^2)
-(1\!+\!\gamma)K \varphi^2  \right]dx
$$
\beq
\ge k_1^2 d^2(\varphi,\psi),          \label{326}
\eeq
where we have used again (\ref{HYP1}) and we have introduced
the constant $k_1^2$
\beq                                                 \label{325}
k_1^2=\min \{\varepsilon^2\omega_3/8, 
\ (2\gamma-1)/4 \}, \qquad \gamma>1/2.
\eeq
Another inequality  of \cite{DacDan98} reads
 \beq                                                 \label{329}
     W_{\gamma}(\varphi,\psi) \leq c_2^2[1+m(d(\varphi,\psi))]
     d^2(\varphi,\psi),
\eeq
where\medskip
\beq
  m(|\varphi |)= \max \{|F_\zeta(\zeta)| \ : \  |\zeta |\leq|\varphi |\}.
                                                      \label{defmfunc}
\eeq
The map $B(d):=[1+m(d)]^{\frac 12}d$ is increasing and continuous, therefore
invertible.
Finally, 
$$
\dot W_{\gamma}(u,u_t) =-\int_0^1 \{\varepsilon u^2_{xx}+
   \varepsilon\gamma u^2_{xt}+a(1\!+\!\gamma)u_t^2+\varepsilon F(u)u_{xx}-
    \varepsilon a u_{xx} u_t  \}dx
$$
\beq
\qquad\:=\!-\!\!\int_0^1 \!\left\{\frac 34 \varepsilon u^2_{xx}\!+ \!
 \varepsilon \left[\frac c2 u_{xx}\!-\!\frac ac u_t\right]^2\!\!+\!
   \varepsilon\gamma u^2_{tx}\!+\!a\!\left[1\!+\!\gamma\!-\!\varepsilon a
   \right]\!u_t^2\!-\!\varepsilon F_u u_x^2 \!\right\}\!dx\label{gigio}
\eeq
\bea
 && \leq  -\int_0^1\{3\varepsilon (1-\lambda)
u_{xx}^2/4 +\varepsilon(3\lambda\pi^2/4-K)u_x^2 \nn
 &&\qquad +[(\varepsilon\pi^2+a)\gamma+a(1-\varepsilon a )]u_t^2\}dx, \nn
\qquad\dot W_{\gamma}(u,u_t) && \leq -\int_0^1\{3\varepsilon (1-\lambda)
\omega_1
(u_{xx}^2+u^2)/4 \!+\!\varepsilon(3\lambda\pi^2/4-K)u_x^2 \label{ciccio}\\
 &&\qquad+[(\varepsilon\pi^2+a)\gamma+a(1-\varepsilon a )]u_t^2\}dx, 
 \nonumber
\eea
where $\lambda\in ]0,1[$ is a constant chosen in such a way that
$3 \lambda\pi^2/4-K>0$, and we have used (\ref{38}), (\ref{HYP1}).

Now we are going to show that for any ``error''
$\sigma>0$ there
exists a $\delta\in]0,\sigma[$ such that
$d(t_0)\equiv d(u_0,u_1)<\delta$ implies
\beq
d(t)\equiv d\left(u(x,t),u_t(x,t)\right)<\sigma       \label{subtesi}
\qquad\forall t\ge t_0.
\eeq
To this end we associate to the neighbourhood with radius
$\sigma$ of the null solution
the Liapunov functional (\ref{321}) choosing the parameter $\gamma$ and 
$\delta$ as the following functions of $\sigma$:
\bea
&& \qquad\gamma(\sigma)=(A\sigma^{\tau}\!+\!A')\varepsilon\!+\!M \qquad\quad
M:=\frac{1\!+\!\varepsilon\pi^2+\!\varepsilon^3\pi^4}{\nu}+
\frac 1{\varepsilon\pi^2}+\frac 12, \label{defgamma}\\
&& \qquad\delta(\sigma)=B^{-1}\left(\frac{\sigma k_1
\left(\gamma(\sigma)\right)}{
c_2\left(\gamma(\sigma)\right)}\right);                  \label{defdelta}
\eea
we shall call the corresponding Liapunov 
functional $W_{\sigma}$. {\it Per absurdum},
assume that there exist a $t_1>t_0$ such that
(\ref{subtesi}) is fulfilled for any $t\in[t_0,t_1[$,
whereas
\beq
d(t_1)=\sigma.                               \label{absurdum}
\eeq
Consider the term in the square bracket at the right-hand side
of (\ref{ciccio}). From (\ref{defgamma}), (\ref{324}), (\ref{HYP3})
considering separately the cases $a>0$, $-\varepsilon\pi^2< a\le 0$,
we find
\beq
\label{argo}
-\left[(\varepsilon\pi^2+a)\gamma+
a(1-\varepsilon a)\right]\leq -1,
\eeq
whence \medskip
\beq
\dot W_{\sigma}(u(t),u_t(t))\leq -k_3^2 d^2(u(t),u_t(t))<0, \label{330}
\eeq
where\medskip
\beq                                                 \label{331}
   k_3^2= \min\{3\varepsilon (1-\lambda)\omega_1/4, \
        \varepsilon(3 \lambda\pi^2/4-K), \  1  \}.
\eeq
From (\ref{326}), (\ref{330}), (\ref{329}), (\ref{defdelta}), it follows
\bea
k_1^2d^2(t_1) &\le& W_{\sigma}(u(t_1),u_t(t_1))<
W_{\sigma}(u(t_0),u_t(t_0))\le c_2^2[1+m(d(t_0))]d^2(t_0) \nn
& < &c_2^2[1+m(\delta)]\delta^2=c_2^2[B(\delta)]^2
=c_2^2\left[B\left(B^{-1}\left(\frac{\sigma k_1}{c_2}\right)\right)
\right]^2=k_1^2\sigma^2, \nonumber
\eea
against (\ref{absurdum}).

Having proved (\ref{subtesi}), it follows $m(d(t))<m(\sigma)$,
which replaced in (\ref{329}) gives
$$                                                 
W_{\sigma} \leq c_2^2(\sigma)[1+m(\sigma)]d^2(t);
$$
together with (\ref{330}) this in turn implies
$$
\dot W_{\sigma}(u(t),u_t(t))\leq  -C(\sigma)W_{\sigma}(u(t),u_t(t)), 
$$
with $C(\sigma):=k_3^2/[c_2^2(\sigma)(1+m(\sigma))]$. Using
the comparison principle we find that $d(t_0)\equiv d(u_0,u_1)<\delta$ implies
\beq
d(u(t),u_t(t))\le D(\sigma)e^{-\frac{C(\sigma)}2 (t-t_0)}d(u_0,u_1),
\eeq
with $D(\sigma):=\frac{c_2}{k_1}\sqrt{1+m(\delta(\sigma))}$.

Last, we show that under
the present assumptions the function (\ref{defdelta}) can be inverted. 
It is evident from (\ref{325}) that
$k_1(\sigma)$ is non-decreasing, from (\ref{35}) and
(\ref{HYP3}) that $\sigma/c_2(\gamma(\sigma)$
is strictly increasing, therefore that 
$(\sigma k_1(\sigma)/c_2(\gamma(\sigma))$ is strictly increasing too,
hence invertible. Since
$B^{-1}$ is invertible, $\delta(\sigma)$ is invertible and its range
is $J$.

Thus we can express $D(\sigma),C(\sigma)$ as functions of $\delta$,
proving the exponential asymptotic stability of the null
solution.
\ep

{\bf Remark 4.} The theorem holds also if we replace the right-hand side of
(\ref{HYP3}) with $A(d)$, where $A:[0,+\infty[\to\b{R}^+$ is any
nondecreasing function such that 
$A(\sigma)/\sigma^2\stackrel{\sigma\to+\infty}{\longrightarrow}0$. 

\bigskip 

{\bf Remark 5.} If (\ref{HYP3}) holds with $\tau=2$ the function
$\frac{\sigma}{c_2\left(\gamma(\sigma)\right)}$ is still increasing
but its range is $[0,2/\varepsilon A]$, implying that
the function
$\frac{\sigma k_1\left(\gamma(\sigma)\right)}{c_2\left(\gamma(\sigma)\right)}$
is still increasing
but its range is $[0,\sqrt{\omega_3}/\sqrt{2} A]$ . Therefore the condition 
(\ref{range}) of definition 3.4 is fulfilled only for 
$\alpha\in]0,B^{-1}(\sqrt{\omega_3}/\sqrt{2} A)[$, and the attraction
region includes the set $d(u_0,u_1)< B^{-1}(\sqrt{\omega_3}/\sqrt{2} A)$.

\bigskip

We now give a variant of the preceding theorem, based on 
a an hypothesis sligtly different from (\ref{HYP3}).
Beside the distance (\ref{33}), we need also a ``weaker''
distance $d_1(u,u_t)$ between the null and a nonnull solution
$u(x,t)$ of the problem  (\ref{31})-(\ref{32}):
for any
$(\varphi,\psi)\in C^2_0([0, 1])\times C_0([0, 1])$
we define
 \beq                                                \label{33bis}
  d_1^2(\varphi,\psi) =
  \int_0^1(\varphi^2+ \varphi_x^2+ \psi^2)dx.
 \eeq
 Clearly,
\beq
d_1(\varphi,\psi) \le d(\varphi,\psi).
\eeq
The ``Hamiltonian'' Liapunov functional $v(u,u_t)$, with
\beq
v(\varphi,\psi):=
\frac{1}{2}\int_0^1 \left\{\psi^2+\varphi_x^2
 -2\left(\int_0^{\varphi(x)}F(z)dz\right)\right\}dx,
\eeq
will play w.r.t. the distance $d_1$ a role similar to
the one played by the Liapunov functionals $V$ or
$W_{\gamma}$ w.r.t. the distance $d$.

\begin{theorem}
Under the following assumptions
\begin{itemize}
\item $F(u)\in C^1(\b{R}), \ F(0)=0$, and
 there exists a positive constant $K$ such that
\beq 
F_u\leq K<3\pi^2/4                                    \label{HYP1bis}
\eeq
\item the function $a$ satisfies
\beq
  \inf a > -\varepsilon\pi^2;      \label{324bis}
\eeq
\item there exists a nondecreasing map $A:J\to J$
such that 
\beq 
|a(x,t,\varphi,\varphi_x,\varphi_{xx},\psi)|
\le A\left[d_1(\varphi,\psi)\right], \label{HYP3bis}
\eeq
\end{itemize}
the null solution of the problem
{\rm (\ref{320})} is exponential-asymptotically stable in the large.
\end{theorem}

\bp{}
Some steps of the proof are exactly as in the previous theorem.
Employing inequality (\ref{zzz}) and the estimate (\ref{38})
we find 
\beq
v \ge  \frac 12 \int_0^1 
\left\{\left(\frac 18 u_x^2+\frac 78 u^2\pi^2\right)+u_t^2-
\frac 34 \pi^2 u^2\right\}dx\nn
\ge \frac 1{16} d_1^2.
\eeq
Setting $v(t)\equiv v(u,u_t)$, integrating by parts
 and using (\ref{320}), (\ref{324bis}),
(\ref{38}) we also find
\bea
\dot v &=& \int_0^1 \left\{u_t[- u_{xx}+u_{tt}-F(u)]\right\}dx
=-\int_0^1 \left\{\varepsilon u^2_{xt}+au_t^2\right\}dx\nn
&\le & -\int_0^1 (\varepsilon \pi^2+a)u_t^2dx<0     \label{ineqdotv}
\eea

Now we are going to prove the uniform boundedness of the solutions
of the problem (\ref{320}).
To this end first note that from the definition (\ref{defmfunc})
it follows
$$
\left\vert\int_0^{\varphi}F(z)dz\right\vert\le m(|\varphi|)\frac{\varphi^2}2;
$$
employing this inequality and the one $\varphi^2\le d_1^2(\varphi,\psi)$
we find
\bea
v &\le&  \frac 12 \left[1+m\left(d_1(u,u_t)\right)\right]d_1^2(u,u_t).
\eea
{} From (\ref{ineqdotv}) we derive the inequality $v(t)<v(t_0)$
for any $t>t_0$, whence
$$
 \frac 1{16}d_1^2(t) \le v(t)<v(t_0)\le \frac 12
\left[1+m\left(d_1(t_0)\right)\right]d_1^2(t_0).
$$
Therefore, for any $t>t_0$
$$
d(t_0)\le \alpha \qquad\Rightarrow \qquad 
d_1(t_0)\le \alpha \qquad\Rightarrow \qquad 
d_1(t)<\beta_1(\alpha):=
2\sqrt{2}\left[1+m(\alpha)\right]^{\frac 12}\alpha,
$$
so that, in view of the assumption (\ref{HYP3bis}), 
\beq
d(t_0)\le \alpha \qquad\Rightarrow \qquad 
|a(x,t,u,u_x,u_t,u_{xx})|\le A\left[\beta_1(\alpha)\right]\equiv
A(\alpha)                               \label{bbla}
\eeq

Now we associate to any $\alpha>0$ 
the Liapunov functional (\ref{321}) with the parameter $\gamma$ 
chosen as the following function of $\alpha$:
\beq
\gamma(\alpha)=A(\alpha)\varepsilon\!+\!M \qquad\qquad
M:=\frac{1\!+\!\varepsilon\pi^2\!+\!\varepsilon^3\pi^4}{\nu}+
\frac 1{\varepsilon\pi^2}+\frac 12;  \label{defgammabis}
\eeq
we shall call the corresponding Liapunov 
functional $W_{\alpha}$. 
Consider the term in the square bracket at the right-hand side
of (\ref{ciccio}). From (\ref{bbla}), (\ref{defgammabis}), 
we find again (\ref{argo}), whence 
\medskip
\beq
\dot W_{\alpha}(u(t),u_t(t))\leq -k_3^2 d^2(u(t),u_t(t))<0, \label{330bis}
\eeq
with the same $k_3^2$ of (\ref{331}).
From (\ref{326}), (\ref{330bis}), (\ref{329}),  it follows
for any $t>t_0$
\bea
k_1^2d^2(t) &\le& W_{\alpha}(u(t),u_t(t))<
W_{\alpha}(u(t_0),u_t(t_0))\le c_2^2[1+m(d(t_0))]d^2(t_0) \nn
& < &c_2^2(\gamma(\alpha))[1+m(\alpha)]\alpha^2=c_2^2(\gamma(\alpha))
B^2(\alpha), \nonumber
\eea
proving the uniform boundedness of $u$:
\beq
d(u(t),u_t(t))< \frac{c_2(\gamma(\alpha))}{k_1(\gamma(\alpha))}B(\alpha) 
\equiv\beta(\alpha).
\eeq

Having proved this, it follows $m(d(t))<m(\beta(\alpha))$,
which replaced in (\ref{329}) gives
$$                                                 
W_{\alpha} \leq c_2^2(\gamma(\alpha))[1+m(\beta(\alpha))]d^2(t);
$$
together with (\ref{330bis}) this in turn implies
$$
\dot W_{\alpha}(u(t),u_t(t))\leq  -C(\alpha)W_{\alpha}(u(t),u_t(t)), 
$$
with $C(\alpha):=k_3^2(\gamma(\alpha))/\{c_2^2(\gamma(\alpha))
[1+m(\beta(\alpha))]\}$. Using
the comparison principle we find that 
$d(t_0)\equiv d(u_0,u_1)\le \alpha$ implies
\beq
d(u(t),u_t(t))\le D(\alpha)e^{-C(\alpha)(t-t_0)}d(u_0,u_1),
\eeq
with $D(\alpha):=\frac{c_2(\gamma(\alpha))}{k_1(\gamma(\alpha))}
\sqrt{1+m(\beta(\alpha))}$, namely the exponential-asymptotical stability.

\ep

\sect{Uniform asymptotic stability in the large for a class 
of non-analytic $f$'s}
\label{nonan}

\indent
~~~~Here we give a generalization of Theorem 2 in  \cite{DanFio00}.
As in the preceding sections, using the trick of the one-parameter
family of Liapunov functionals we are able to
replace the boundedness assumption 
for the function $a$ by a weaker one.

\begin{theorem}
Under the following assumptions
\bea
&&  F(\varphi)\in C(\b{R})\mbox{ such that } F(0)=0, \\
&&  \mbox{there exist $\tau\in[0,1[$ and $D>0$ such that,
for any $\varphi,\psi$} \label{HYP1'} \\
&& 0 \le -\int_0^1\left(\int_0^{\varphi(x)}F(z)dz \right) dx
\le D d^{\tau+1}(\varphi,\psi), \nn
&&
\int_0^1 F(\varphi(x)) \varphi_{xx}(x)dx \ge 0\ \
 \mbox{for any }\varphi\in C^2_0([0,1]),
 \label{HYP3'}\\
&& \mbox{the function $a$ satisfies }\:\:
  \inf a > -\varepsilon\pi^2, \label{324'}\\
&& \mbox{there exists a nondecreasing map $A:[0,\infty[\to\b{R}^+$
such that} \label{HYPF}\\
&&|a(x,t,\varphi,\varphi_x,\varphi_{xx})|\le A\left(d(\varphi,\psi)\right), 
                                                   \nonumber
\eea
the null solution of the problem
{\rm (\ref{320})} is uniformly asymptotically stable in the large.
\end{theorem}

\bp{} From (\ref{lal}), (\ref{HYP1'})
\bea
W_{\gamma}(\varphi,\psi) &\ge &
\frac{1}{2}\int_0^1  \{(\gamma-1/2)\psi^2+(1+\gamma)
\varphi_x^2+\varepsilon^2 \varphi_{xx}^2/4  \}dx \nn
&\ge&\frac{1}{2}\int_0^1  \{(\gamma-1/2)\psi^2+(1+\gamma)\omega_3
(\varphi^2+\varphi^2_x)+\varepsilon^2 \varphi_{xx}^2/4  \}dx \nn
&\ge& k_1'{}^2 d^2(\varphi,\psi),                             \label{lalla}
\eea
where
\beq
 k_1'{}^2:=\frac 12\min\left\{\gamma-\frac 12,\frac{\varepsilon^2}4,
(1+\gamma)\omega_3 \right\}, \qquad\qquad \gamma>\frac 12.
\eeq
Moreover, taking into account (\ref{321}), assumption (\ref{HYP1'}), 
 noting that
$(\varepsilon \varphi_{xx}-\psi)^2\le\varepsilon^2 \varphi_{xx}^2+
\psi^2+\varepsilon(\varphi_{xx}^2+\psi^2)$, and considering
(\ref{35}) it follows
\beq                                                 \label{328}
W_{\gamma}(\varphi,\psi) \leq G_{\gamma}\left(d(\varphi,\psi)\right),
\eeq
where
\beq
G_{\gamma}(d):=c_2^2(\gamma)d^2+D(\gamma+1) d^{\tau+1}.  \label{defGamma}
\eeq
For any choice of $\gamma> \frac 12$
the map $G_{\gamma}(d)$ is increasing and continuous in $d$,
therefore invertible.
Finally, with the help
of (\ref{38}) we obtain from (\ref{gigio})
\bea
\dot W_{\gamma}(u,u_t) &\leq& -\int_0^1 \{(3/4)\varepsilon u^2_{xx}+
[\varepsilon\gamma+a(1+\gamma-\varepsilon a)]u_t^2\}dx \nn
&\le & -\int_0^1 \{\varepsilon\omega_2 (u^2_{xx}+u^2_x+u^2)/4+
[(\varepsilon+a)\gamma+a(1-\varepsilon a)]u_t^2\}dx. \label{qua}
\eea

Now we are going to show that for any ``error''
$\sigma>0$ there
exists a $\delta\in]0,\sigma[$ such that
$d(t_0)\equiv d(u_0,u_1)<\delta$ implies
\beq
d(t)\equiv d\left(u(x,t),u_t(x,t)\right)<\sigma       \label{subtesi'}
\qquad\forall t\ge t_0.
\eeq
To this end we choose the parameter $\gamma$ in
the Liapunov functional (\ref{321})  as in (\ref{defgammabis}) and 
$\delta$ as the following function of the error $\sigma$:
\beq
\delta(\sigma)=G_{\gamma(\sigma)}^{-1}
\left( \sigma^2 k_1'^2\left(\gamma(\sigma)\right)\right);   \label{defdelta'}
\eeq
we shall indicate the corresponding Liapunov 
functional $W_{\gamma(\sigma)}$ simply by $W_{\sigma}$. {\it Per absurdum},
assume that there exist a $t_1>t_0$ such that
(\ref{subtesi}) is fulfilled for any $t\in[t_0,t_1[$,
whereas (\ref{absurdum}) holds for $t=t_1$.
Consider the term in the square bracket at the right-hand side
of (\ref{qua}). From (\ref{defgammabis}), (\ref{324}), (\ref{HYPF})
we get again (\ref{argo}), whence
\beq
\dot W_{\sigma}(u(t),u_t(t))\leq -k_3'{}^2 d^2(u(t),u_t(t))<0, \label{330'}
\eeq
where now $k_3'{}^2:=\min\{\varepsilon\omega_2/4,1\}$.
From (\ref{lalla}), (\ref{328}), (\ref{330'}),(\ref{defdelta'}), it follows
\bea
k_1'^2d^2(t_1) &\le& W_{\sigma}(u(t_1),u_t(t_1))<
W_{\sigma}(u(t_0),u_t(t_0)) \nn
&\le &G_{\gamma(\sigma)}(d(t_0))
 < G_{\gamma(\sigma)}(\delta(\sigma))=k_1'^2\sigma^2, \nonumber
\eea
against (\ref{absurdum}).
So we have proved the uniform stability of the null solution.

Note now 
that the function $\delta(\sigma)$ is invertible, since it is the
composition of two increasing functions. Therefore $W_{\sigma}$ 
can be expressed
as a function $W_{\delta}$ of the parameter $\delta$. By (\ref{330'})
it is $W_{\delta}(t)\le W_{\delta}(t_0)$ so by
(\ref{lalla}), (\ref{328}) we find that for $d(t_0)\equiv d(u_0,u_1)\le\delta$
$$
d^2(t)\le\frac{W_{\delta}(t)}{k_1'^2}\le\frac{W_{\delta}(t_0)}{k_1'^2}\le
\frac{G_{\gamma(d(t_0))}}{k_1'^2}\le
\frac{G_{\gamma(\delta)}}{k_1'^2\big(\gamma(\delta)\big)}=:\beta^2(\delta),
$$
proving the uniform boundedness of $u$.

Employing an argument of  \cite{DanFio00} one can
now show that for any choice of the
initial condition $d(t_0)<\delta$ 
the functional $W_{\delta}$ decreases to zero (at least) 
as a negative power
of $(t-t_0)$ as $(t-t_0)\to \infty$. 
From (\ref{328}) we find
$$
d^2\ge \min\left\{\frac {W_{\delta}}{2c_2^2(\gamma(\sigma))}, 
\left(\frac {W_{\delta}}{2D(\gamma\!+\!1)}\right)^{\frac 2{\tau+1}}
\right\},
$$
which considered in (\ref{330'}) gives
\beq
\dot W_{\delta}(u,u_t)\le -k_3' \min\left\{\frac{W_{\delta}}{2c_2^2},
\left(\frac {W_{\delta}}{2D(\gamma\!+\!1)}\right)^{\frac 2{\tau+1}}
\right\}\le 0.                                      \label{wpp}
\eeq
If at $t=t_0$
\beq
\frac {W_{\delta}}{2c_2^2} \ge \left(\frac {W_{\delta}}
{2D(\gamma\!+\!1)}\right)^{\frac 2{\tau+1}},           \label{situation1}
\eeq
then setting 
$E(\delta):=\frac {k_3'}{\left[2D\big(\gamma
(\delta)\!+\!1\big)\right]^{\frac 2{\tau+1}}}\frac{1-\tau}{1+\tau}>0$ one finds
\beq
d^2(t) \le \frac{W_{\delta}(t)}{k_1'^2}\le
\frac 1{k_1'^2[W_{\delta}(t_0)+E(t-t_0)]^{\frac{1+\tau}{1-\tau}}}\le
\frac 1{k_1'^2[E(t-t_0)]^{\frac{1+\tau}{1-\tau}}}              \label{final1}
\eeq
for $t\ge t_0$. If on the contrary
\[
\frac {W_{\delta}(t_0)}{2c_2^2} < 
\left(\frac {W_{\delta}(t_0)}{2D(\gamma\!+\!1)}\right)^{\frac 2{\tau+1}},
\]
(\ref{wpp}) will imply for some time
\[
\dot W_{\delta}(u,u_t)\le -k_3'  W_{\delta}
\]
and by the comparison principle an (at least) exponential decrease
of $W_{\delta}$.  Hence there will
exist a $\tilde T(\delta)>0$ such that
\[
\frac{W_{\delta}(t_0+\tilde T)}{2c_2^2} = 
\left(\frac {W_{\delta}(t_0+\tilde T)}{2D(\gamma\!+\!1)}\right)^{\frac 2{\tau+1}},
\]
after which (\ref{wpp}) will take again the form considered in
the previous case and thus imply
\beq
d^2(t) \le \frac{W_{\delta}(t)}{k_1'}
\le \frac 1{k_1'^2[W_{\delta}(t_0+\tilde T)+E(t-t_0-
\tilde T)]^{\frac{1+\tau}{1-\tau}}}
\le \frac 1{k_1'^2[E(t-t_0-
\tilde T)]^{\frac{1+\tau}{1-\tau}}}       \label{final2}
\eeq
for $t\ge t_0+\tilde T$. Formula (\ref{final2})  will be valid also
if $\delta$ is so small that inequality (\ref{situation1}) occurs, 
provided we correspondingly 
define $\tilde T:=0$, so that it reduces to (\ref{final1}). 
Formula (\ref{final2}) evidently implies the quasi-uniform asymptotic 
stability in the large of the null solution. \ep


\begin{thebibliography}{99}

\bibitem{cannon} J. R. Cannon, {\it The One-Dimensional Heat Equation},
      Addison-Wesley (1984).

\bibitem{DacDan98}  B. D'Acunto, A. D'Anna, {\it Stability for a 
third order Sine-Gordon equation}, Rend. Mat. Serie {\bf VII}, Vol. 18, 
p. 347 (1998).

\bibitem {dr1} B. D'Acunto, P. Renno, {\it On Some Nonlinear Visco-elastic
       Models}, Ricerche di Mat., {\bf 41}, p. 101 (1992).

\bibitem {dr2} B. D'Acunto, P. Renno, {\it  On the operator
        $ \varepsilon \partial_{xxt}+c^2\partial_{xx}-\partial_{tt} $
        in General Domains}, Atti del Seminario Matematico e Fisico
        dell'universit\'a di Modena, Vol. {\bf XLVII} (1999), 191.

\bibitem{DanFio00} A. D'Anna, G. Fiore
{\it Stability and attractivity for a class of
dissipative phenomena}, Rend. Mat. Serie {\bf VII}, Vol. 21, 
p. 191 (2000).

 \bibitem{dav} A. S. Davydov, {\it Solitons in Molecular Systems}, Reidel
 Publishing Company (1985).

 \bibitem{ghi} J. M. Ghidaglia, A. Marzocchi, {\it Long time behaviour of
 strongly damped wave equations, global attractors and their dimensions},
 SIAM, {\bf 22}, p. 879 (1990).

 \bibitem{gr1} J. M. Greenberg, R. C. MacCamy and V. J. Mizel, {\it On the
 Existence, Uniqueness, and Stability of Solutions of the Equation
 \( \sigma'(u_x)u_{xx}+\lambda u_{xtx}=\rho_0u_{tt} \)}, J. Math. Mech.,
 {\bf 17} (7), p. 707 (1969).

 \bibitem{gr2} J. M. Greenberg and R. C. MacCamy, {\it On the Exponential
 Stability of Solutions of \( E(u_x)u_{xx}+\lambda u_{xtx}=\rho u_{tt} \)},
 J. Math. Analysis Appl., {\bf 31}, p. 406 (1970).

 \bibitem{Hale} J. K. Hale, {\it Asymptotic behaviour of solutions of 
 differential-difference equations}, Proc. of Symp. Nonlinear Oscillations,
 IUTAM, Kiev, September 1961, II, 409-426.
 
 \bibitem{lsc} P. S. Lomdhal, O. H. Soerensen, P. L. Christiansen,
 {\it Soliton Excitations in Josephson Tunnel Junctions}, Phys. Rev. B
 {\bf 25},  p. 5337 (1982).

 \bibitem{mor} J. A. Morrison, {\it Wave propagations in rods of Voigt
 material and visco-elastic materials with three-parameters models}, Quart.
 Appl.  Math., {\bf 14}, p. 153 (1956).

\bibitem{rio} N. Flavin, S. Rionero, {\it Qualitative Estimates for Partial
          Differential Equations}, CRC Press, (1996).

\bibitem{yos} T. Yoshizawa, {\it Stability Theory by Liapunov's second
 method}, The Mathematical Society of Japan, (1966).

\end{thebibliography}
\end{document}